\documentclass[aoas]{imsart}

\usepackage{tablefootnote}
\usepackage{amsthm}
\usepackage{amsmath}
\usepackage{amssymb}
\usepackage{amsfonts}
\usepackage{mathrsfs}
\usepackage{graphicx}
\usepackage{colonequals}
\usepackage[dvipsnames]{xcolor}
\usepackage[authoryear,sort&compress]{natbib}
\usepackage{hyperref}
\usepackage{bm}
\usepackage{}
\usepackage[noline]{algorithm2e}
\usepackage{booktabs}
\usepackage{multirow}

\hypersetup{
colorlinks=true,
linkcolor=black,
filecolor=black,
citecolor=black,
urlcolor=NavyBlue
}

\startlocaldefs
\theoremstyle{plain}

\newtheorem{theorem}{Theorem}[section]

\theoremstyle{remark}
\newtheorem{definition}[theorem]{Definition}

\newcommand{\bs}[1]{\boldsymbol{#1}}

\def\y{\bs{y}}

\def\x{\bs{x}}

\def\b{\bs{b}}

\def\X{\bs{X}}

\def\bga{\bs{\gamma}}
\def\balpha{\bs{\alpha}}
\def\bmu{\bs{\mu}}
\def\bsig{\bs{\sigma}}
\def\bdel{\bs{\delta}}
\def\bbar{\bar{\b}}

\def\bpi{\bs{\pi}}

\def\BF{\mathrm{BF}}

\DeclareMathOperator*{\argmax}{argmax}

\endlocaldefs

\begin{document}

\begin{frontmatter}

\title{Bayesian variable selection in a Cox proportional hazards
  model with the ``Sum of Single Effects'' prior}
\runtitle{Variable selection in a CoxPH model with the SuSiE prior}
\runauthor{Y. Yang, K. Tayeb, P. Carbonetto and M. Stephens}

\begin{aug}
\author[A]{\fnms{Yunqi}~\snm{Yang}\ead[label=e1]{yunqiyang@uchicago.edu}},
\author[A]{\fnms{Karl}~\snm{Tayeb}\ead[label=e2]{ktayeb@uchicago.edu}},
\author[B]{\fnms{Peter}~\snm{Carbonetto}\ead[label=e3]{pcarbo@uchicago.edu}}
\author[B]{\fnms{Xiaoyuan}~\snm{Zhong}\ead[label=e4]{xzhong999@uchicago.edu}}
\author[B]{\fnms{Carole}~\snm{Ober}\ead[label=e5]{cober@uchicago.edu}}
\and
\author[C]{\fnms{Matthew}~\snm{Stephens}\ead[label=e6]{mstephens@uchicago.edu}}
\address[A]{Committee on Genetics, Genomics and System Biology,
    University of Chicago, Chicago, IL\printead[presep={,\ }]{e1,e2}}
\address[B]{Department of Human Genetics,
    University of Chicago, Chicago, IL\printead[presep={,\ }]{e3,e4,e5}}
\address[C]{Departments of Statistics and Human Genetics,
    University of Chicago, Chicago, IL\printead[presep={,\ }]{e6}}
\end{aug}

\begin{abstract}
Motivated by genetic fine-mapping applications, we introduce a new
approach to Bayesian variable selection regression (BVSR) for
time-to-event (TTE) outcomes.  This new approach is designed to deal
with the specific challenges that arise in genetic fine-mapping,
including: the presence of very strong correlations among the
covariates, often exceeding 0.99; very large data sets containing
potentially thousands of covariates and hundreds of thousands of
samples. We accomplish this by extending the ``Sum of Single Effects''
(SuSiE) method to the Cox proportional hazards (CoxPH) model. We
demonstrate the benefits of the new method, ``CoxPH-SuSiE'', over
existing BVSR methods for TTE outcomes in simulated fine-mapping data
sets. We also illustrate CoxPH-SuSiE on real data by fine-mapping
asthma loci using data from UK Biobank. This fine-mapping identified
14 asthma risk SNPs in 8 asthma risk loci, among which 6 had strong
evidence---a posterior inclusion probability greater than 50\%---for
being causal. Two of the 6 putatively causal variants are known to be
pathogenic, and others lie within a genomic sequence that is
known to regulate the expression of {\em GATA3}.
%
%
\end{abstract}

\begin{keyword}
\kwd{time-to-event data}
\kwd{survival analysis}
\kwd{Cox proportional hazards model}
\kwd{Bayesian variable selection in regression}
\kwd{genomics}
\kwd{genome-wide association studies}
\kwd{genetic fine-mapping}
\kwd{UK Biobank}
\end{keyword}

\end{frontmatter}

\renewcommand\figurename{Figure}

\section{Introduction}

With the increasing availability of biobanks and electronic health
records, analyzing time-to-event (TTE) phenotypes such as disease age
of onset
%
%
has become more common in genetics. TTE data can yield insight into
the genetics of disease development and progression, and enhance our
understanding of disease etiology.
%
%
Research has shown that modeling TTE phenotypes using survival models
can be more powerful than modeling binary disease status using
logistic models,
%
%
particularly for events that are more common \citep{green1983comparison,
  callas1998empirical, staley2017comparison, hughey2019cox}. 
%
%
Additionally, genome-wide association studies (GWAS) of TTE
phenotypes have identified significant loci
%
%
that are not significant based on case-control status
\citep{bi2020fast}.

After detecting a genetic association, a common next step is
``fine-mapping'', which is the process of narrowing the set of
associated genetic variants down to a smaller set of causal candidates
\citep{hutchinson-2020-review, maller2012bayesian, kote2013fine,
  schaid2018from}. The most widely used fine-mapping approaches frame
the fine-mapping as a variable selection problem within a regression
model, in which the variables in the regression are the genetic
variants \citep{george-mcculloch-1997, raftery-1997,
  sillanpaa2005bayesian}.  Fine-mapping is a particularly challenging
variable selection problem because of the strong correlations that can
exist among genetic variants; pairs of nearby genetic variants can
have sample correlations of 0.99 or higher.
%
%

The very strong correlations, or ``linkage disequilibrium'' (LD),
among the genetic variants often means that we cannot select variables
confidently. For this reason, quantifying uncertainty in variable
selection becomes important, and Bayesian approaches to variable
selection in regression (BVSR) are particularly good at dealing with
this. BVSR methods for fine-mapping typically provide two key
quantities: a posterior inclusion probability (PIP) for each variable,
which gives the probability that this variable is a causal variable
(which we use as shorthand for a variable with a truly non-zero
regression coefficient); and ``Credible Sets'' (CSs) of variables,
which are defined as sets of variables that have a high probability
(e.g., 0.95 or greater) of containing at least one causal variable
\citep{maller2012bayesian, wang2020simple}.


Several BVSR methods and software for fine-mapping quantitative traits
are available, including \cite{hormozdiari2014identifyingassociation,
  kichaev2014integrating, maller2012bayesian, yang2012conditional,
  chen2015fine, riviera, wen2016efficient, wallace2015dissection,
  wang2020simple, benner2016finemap:studies, lee2018bayesian,
  zou2022fine}. All these methods are based on linear regression
models with Gaussian errors. For analyzing TTE phenotypes using
survival models, however, the options are limited, and none of the
available methods provide CSs.
%
%
\cite{newcombe2017weibull} propose methods for sparse Bayesian Weibull
regression with a normal prior for the non-zero effect variables, and
they performed posterior inferences using a reversible jump Markov
chain Monte Carlo (MCMC) algorithm. \cite{nikooienejad2020bayesian}
developed a BVSR method using the Cox proportional hazards (CoxPH) 
model \citep{cox1972regression},
%
%
which is the most widely used model for TTE outcomes. They used a
nonlocal prior for the non-zero regression coefficients and a
stochastic search algorithm for inference. More recently,
\cite{komodromos2022variational} introduced a sparse variational Bayes
approach for the CoxPH model, with a Laplace prior for the non-zero
effect variables.
%
%
While all of these methods provide PIPs, none of them provide the CSs.
%
%
Further, some of these methods have not been tested on data sets
containing strongly correlated variables,
leaving open the question of how well they will perform in
fine-mapping settings with very strongly correlated genetic variants.

Here we introduce a new BVSR method for TTE outcomes that is designed
to address these issues: it provides both CSs and PIPs, and it is
specifically designed to cope with highly correlated covariates. This
new method combines the CoxPH model with the ``Sum of Single Effects''
(SuSiE) method introduced in \cite{wang2020simple}. SuSiE was
specifically designed to provide CSs and PIPs, and to deal with the
very high correlations that occur in fine-mapping applications. It is
also fast, and able to deal with large data sets containing
thousands of covariates.
%
%
Our new fine-mapping method for TTE outcomes, ``CoxPH-SuSiE'', extends the
advantages of SuSiE to the CoxPH regression model. Further, this
approach of combining SuSiE with a single-variable regression model is
quite general, and could be easily adapted to other regression models
such as logistic regression or Poisson regression. R code implementing
the new methods is available online at
\url{https://github.com/yunqiyang0215/survival-susie/}
(see also \citealt{zenodo}).

\subsection{Organization of the paper}

The remainder of the paper is organized as follows. Section
\ref{sec:background} presents necessary background. Section
\ref{sec:coxph-susie} describes our new approach, CoxPH-SuSiE.
Section \ref{sec:BFcomp} numerically compares the accuracy of
different approaches for computing the Bayes factors for CoxPH
regression models, which is a key step in CoxPH-SuSiE. Then Section
\ref{sec:simulations} assesses the performance of CoxPH-SuSiE and
other BVSR methods in simulated fine-mapping data sets with a TTE
outcome. Section \ref{sec:coxph_susie_real_data} presents an
application of CoxPH-SuSiE to fine-mapping asthma loci using data from
UK Biobank. Finally, Section \ref{sec:coxph_susie_discussion}
discusses benefits and practical limitations of CoxPH-SuSiE, and
possible future directions.

\section{The CoxPH and SuSiE models}
\label{sec:background}

Our new methods involve combining two essential building blocks: the Cox proportional hazards (CoxPH) model for survival analysis \citep{cox1972regression}, and the SuSiE prior model for BVSR \citep{wang2020simple}, which we now describe.


\subsection{The CoxPH model}
\label{sec:survival}

The CoxPH model is a model for time-to-event (TTE) data in a
prospective cohort study where the measured data on each individual
$i$ is {\em either} the time $T_i$ at which a particular event
(e.g., disease diagnosis) occurs {\em or} the time $C_i$ at which
the individual left the study without experiencing the event, known as
the ``censoring'' time. For notational convenience we assume that
every individual has both an event time and a censoring time, but we
only observe whichever occurs first.  Thus the outcomes for a cohort
of $n$ individuals are denoted as $(\y, \bdel)$, where $\y = (y_1,
\ldots, y_n)$, $\bdel = (\delta_1, \ldots, \delta_n)$, $y_i =
\min(T_i, C_i)$, and $\delta_i = \mathbb{I}(T_i \le C_i)$ is 1 when
individual $i$ was uncensored, and zero otherwise.  The total number of
observed events is $K \colonequals \sum_{i=1}^n \delta_i$, and the
ratio $\frac{n-K}{n}$ is referred to as the ``censoring rate''.

CoxPH regression models the relationship between the event times $T_i$
and covariates, $\x_i = (x_{i1}, \ldots, x_{ip})^{\intercal}$, through
the {\em hazard function} $\lambda$, defined as
\begin{equation}
\lambda_i(t) =
\lim_{\Delta \,\to\, 0} \frac{1}{\Delta} \Pr(t \leq T < t + \Delta \mid
T \ge t, x_i)
\label{eq:hazard}.
\end{equation}
Specifically, CoxPH assumes
\begin{equation}
\lambda_i(t) = \lambda_0(t)\exp(\b^{\intercal}\x_i),
\end{equation}
where $\b = (b_1, \ldots, b_p)^{\intercal}$ is a vector of
coefficients to be estimated, and $\lambda_0(t)$ is some (unknown)
baseline hazard function. \cite{cox1972regression} proposed to
estimate $\b$ by maximizing a conditional likelihood, also known as
the {\em partial likelihood}. The partial likelihood has the advantage
that it does not depend on the unknown baseline hazard,
$\lambda_0(t)$. Additionally, the partial likelihood can deal with
censoring \citep{cox1975partial}.
%
%

\subsection{The SuSiE prior}
\label{sec:susie}

The SuSiE model was introduced in \cite{wang2020simple} in the context
of linear regression. The SuSiE model combines the linear regression
likelihood with a prior on the regression coefficients $\b$ that we
call the ``SuSiE prior''.  The SuSiE prior constructs the regression
coefficients $\b$ as the sum of $L$ ``single effect vectors'', $\b_l$,
each of which has exactly one non-zero element, whose value is
$b_l$:
\begin{equation}
\begin{aligned}
\b &= \sum_{l=1}^L \b_l \\
\b_l &= b_l \bga_l \\
b_l&\sim N(0,\sigma_{0l}^2) \\
\bga_l&\sim \text{Multinom}(1,\bpi),
\end{aligned}
\end{equation}
where $N(\mu, \sigma^2)$ denotes the univariate normal distribution
with mean $\mu$ and variance $\sigma^2$; $\text{Multinom}(n,\bpi)$
denotes the multinomial distribution for $n$ multinomial trials with
probabilities $\bpi = (\pi_1, \ldots, \pi_p)$; and $\bga_l \in \{0,
1\}^p$ is a binary vector indicating which element of $\b_l$ is
non-zero. Like \cite{wang2020simple}, we assume $\pi_1 = \dots = \pi_p
= \frac{1}{p}$, so all covariates are equally likely to have non-zero
coefficients, although other values for $\bpi$ could be used.

Because each $\b_l$ has exactly one non-zero entry, the SuSiE prior
ensures that $\b$ has at most $L$ non-zeros.  In fine-mapping
settings, $L$ is typically small (e.g., $L=10$), and so the SuSiE
prior is sparse.  For now, we assume $L$ and the prior variances
$\sigma^2_{0l}$ are pre-specified; later, we discuss estimating these
hyperparameters.

\section{The CoxPH-SuSiE model}
\label{sec:coxph-susie}

CoxPH-SuSiE performs Bayesian variable selection for survival analysis
by combining the CoxPH partial likelihood (Section~\ref{sec:survival})
with the SuSiE prior (Section~\ref{sec:susie}).  Like all other
Bayesian variable selection methods, the inferences in CoxPH-SuSiE are
based on Bayes' Theorem, except that the likelihood in Bayes' Theorem
is replaced with the partial likelihood.  \cite{kalbfleisch1978non}
showed that using the partial likelihood implicitly approximates
standard Bayesian inferences by assuming a limiting gamma process
prior for the baseline hazard $\lambda_0(t)$; see also
\cite{sinha2003bayesian, ibrahim2001bayesian}. We refer to the model
that combines the CoxPH partial likelihood with the SuSiE prior as the
``CoxPH-SuSiE model''.

In the following sections, we derive the core underlying posterior
computations for CoxPH-SuSiE as well as an efficient algorithm for
fitting the CoxPH-SuSiE model.

\subsection{Bayesian CoxPH regression for a single variable}
\label{sec:bayes_simple}

To build toward the CoxPH-SuSiE model, we first consider a much
simpler CoxPH regression model with a {\em single} covariate, $\x =
(x_1, \dots, x_n)^{\intercal}$, and a {\em single} regression
coefficient, $b$. In this model, we assume a normal prior for $b$, and
for later use we allow for a fixed offset, ${\bm c} = (c_1, \dots,
c_n)^{\intercal}$:
\begin{equation}
\begin{aligned}
\lambda_i(t) &= \lambda_0(t)\exp(bx_i + c_i) \\
b&\sim N(0, \sigma_0^2).
\end{aligned}
\label{eq:coxph}
\end{equation}
We call this the ``single-variable CoxPH regression model''.

Let $\ell(b; \x, {\bm c})$ denote the Cox partial likelihood for this
model. (This likelihood, as well as other quantities, also depend on
$\y, \bdel$, but to keep notation light we suppress
this dependence.) Let $\hat{b}(\x, {\bm c}) \colonequals
\argmax_b \ell(b; \x, {\bm c})$ denote the maximum partial likelihood
estimate, and $s(\x, {\bm c})$ denote its (estimated) standard error.
For notational simplicity we often write $\hat{b}$ and $s$,
suppressing their explicit dependence on $\x$ and ${\bm c}$.  Both
$\hat{b}$ and $s$ are easily obtained from standard software, such
as the {\tt coxph()} function in the survival R package
\citep{survival-book}.
 
The posterior distribution for $b$ under this model is
\begin{equation}
\label{eq:post}
p(b \mid \x, \y, \bdel, {\bm c}, \sigma_0^2) \propto
\ell(b; \x, {\bm c}) \, p(b \mid \sigma_0^2),
\end{equation}
and the Bayes Factor (BF) \citep{bayes_factors} comparing this model
to the null model ($b = 0$) is
\begin{equation}
\label{eq:BFp}
\BF(\x, {\bm c}, \sigma_0^2) \colonequals
\frac{\int \ell(b;  \x, {\bm c}) \, p(b \mid \sigma_0^2) \,db}
     {\ell(0; \x, {\bm c})}.
\end{equation}
(The null likelihood $\ell(0; \x, {\bm c})$ does not depend on $\x$,
but $\x$ is included in the notation for consistency.)

Neither the posterior distribution \eqref{eq:post} nor the BF
\eqref{eq:BFp} have a closed-form solution. While it would be possible
to approximate these quantities using numerical integration methods,
here we take a simpler (and potentially faster) approach by
taking a quadratic approximation to the partial
log-likelihood,
\begin{equation}
\log \ell(b; \x, {\bm c}) \approx
\log \ell(\hat b; \x, {\bm c})
- \frac{(b - \hat{b})^2}{2s^2},
\end{equation}
which yields the following approximation to the likelihood:
\begin{equation}
\ell(b; \x, {\bm c}) \approx
\hat{\ell}(b; \x, {\bm c}) \colonequals
\ell(\hat b ; \x, {\bm c})
\textstyle \exp\{-{\frac{1}{2s^2}}(b - \hat{b})^2\}.
\end{equation}
This approximation yields a Gaussian posterior distribution for $b$,
\begin{equation} 
b \mid  \x, \y, \bdel, {\bm c}, \sigma_0^2 \sim N(\mu_1, \sigma_1^2),
\end{equation}
where
\begin{align}
\sigma_1^2(\x, {\bm c},\sigma_0^2) \colonequals&\;
\frac{1}{1/s^2 + 1/\sigma_0^2}
\label{eq:sigma1} \\
\mu_1(\x, {\bm c},\sigma_0^2) \colonequals&\;
\frac{\sigma_1^2(\x, {\bm c}, \sigma_0^2)}{s^2} \times \hat{b}.
\label{eq:mu1}
\end{align}
Letting $z \colonequals \hat{b}/s$, the corresponding approximate BF is
\begin{align}
\widehat\BF(\x, {\bm c}, \sigma_0^2) 
\colonequals&\;
\frac{\int \hat \ell(b; \x, {\bm c}) \, p(b \mid \sigma_0^2) \, db}
     {\ell(0; \x, {\bm c})} \nonumber \\
=&\; \mathrm{ABF}(\x, {\bm c}, \sigma_0^2)
\times \exp\bigg(\frac{-z^2}{2}\bigg) \times
\frac{\ell(\hat b; {\bm x}, {\bm c})}{\ell(0; {\bm x}, {\bm c})},
\label{eq:BF_lap}
\end{align}
where $\mathrm{ABF}(\x, {\bm c}, \sigma_0^2)$ denotes the
``asymptotic Bayes factor'' \citep{wakefield2009bayes},
\begin{equation}
\mathrm{ABF}(\x, {\bm c}, \sigma_0^2) =
\sqrt{\frac{s^2}{\sigma_0^2 + s^2}}
\times \exp\bigg(\frac{z^2}{2} \times
                 \frac{\sigma_0^2}{\sigma_0^2 + s^2}\bigg).
\label{eq:abf}
\end{equation}
The approximation \eqref{eq:BF_lap} is a variant of the standard
Laplace approximation to the Bayes Factor \citep{bayes_factors}.
Comparisons below (Section \ref{sec:BFcomp}) show that this Laplace
approximation to the Bayes factor (``Laplace BF'') is accurate, and
can be substantially more accurate than the ABF approximation
\eqref{eq:abf}.



\subsection{Single effect regression for the Bayesian CoxPH model}
\label{sec:coxph_ser}

\cite{wang2020simple} defined a ``single effect regression'' (SER) as
a multiple regression model in which {\em exactly one of $p$
  covariates} has a non-zero coefficient in the model. (The SER model
is a special case of the SuSiE model, with $L = 1$.) The analogous SER
model for CoxPH regression with fixed offset ${\bm c}$ is
\begin{equation}
\begin{aligned}
\lambda_i(t) &= \lambda_0(t)
\exp(\b^{\intercal}\x_i + c_i) \\
\b &= b\bga \\
\bga &\sim\text{Multinom}(1,\bpi)\\
b &\sim N(0, \sigma_0^2).
\end{aligned}
\label{eq:coxph-ser-model}
\end{equation}
In the following, we assume the data for the $p$ covariates are stored
as an $n \times p$ matrix $\X$, in which ${\bm x}_i$ denotes the $i$th
row of $\X$ and ${\bm x}_{\cdot j}$ denotes the $j$th column of $\X$.
%
%

Under this model, the posterior distribution of $\bga$ is
\begin{equation}
\begin{aligned}
& \bga \mid \X, \y, \bdel \sim \text{Multinom}(1,\balpha) \\ 
&  \Pr(\bga_j = 1 \mid \X, \y, \bdel) =
\frac{\pi_j\BF_j}{\sum_{j'=1}^p \pi_{j'} \BF_{j'}},
\end{aligned}
\end{equation}
where $\BF_j \colonequals \BF( \x_{\cdot j}, {\bm c},\sigma_0^2)$ is
the Bayes factor \eqref{eq:BFp} that compares the CoxPH model with the
$j$th column of $\X$ as the covariate vs. the null model.  In
practice, we compute these posterior probabilities by replacing the
exact Bayes factors with their approximations,
\begin{equation} \label{eq:alpha}
\Pr(\bga_j = 1 \mid \X, \y, \bdel) \approx \alpha_j \colonequals
\frac{\pi_j\widehat\BF_j}{\sum_{j'=1}^p \pi_{j'} \widehat\BF_{j'}},
\end{equation}
where $\widehat\BF_j \colonequals \widehat\BF(\x_{\cdot j}, {\bm
  c},\sigma_0^2)$; see eq.\eqref{eq:BF_lap}. The corresponding
approximate posterior distribution of $b$ given $\bga_j = 1$ is
normal with mean $\mu_{1j}$ and variance $\sigma_{1j}^2$, 
where 
\begin{align}
\sigma^2_{1j} &= \sigma_1^2(\x_{\cdot j}, {\bm c}, \sigma_0^2)
\label{eq:sigma1j} \\
\mu_{1j} &= \mu_1(\x_{\cdot j}, {\bm c}, \sigma_0^2),
\label{eq:mu1j}
\end{align}
using the definitions in (\ref{eq:sigma1}, \ref{eq:mu1}).  

In summary, computing the approximate posterior distribution for the
CoxPH SER model on $p$ covariates involves computing three
$p$-vectors, $\balpha = (\alpha_1, \ldots, \alpha_p)$, $\bmu_1 =
(\mu_{11}, \ldots, \mu_{1p})$ and $\bsig_1^2 = (\sigma_{11}^2, \ldots,
\sigma_{1p}^2)$, whose elements are given by
(\ref{eq:alpha}--\ref{eq:mu1j}). To express these posterior
computations succinctly, we define the following function:
\begin{align}
\label{eq:coxph-ser}
\text{\sc CoxPH-SER}(\X, \y, \bdel, {\bm c}; \sigma_0^2)
\colonequals (\balpha, \bmu_1, \bsig_1^2).
\end{align}
We note that, given these quantities, the posterior mean of the single
effect vector $\b$ is easily computed as $\bar{\b} = \balpha \circ
\bmu_{1}$ where ``$\circ$'' denotes element-wise multiplication.

\paragraph*{Estimating the prior variance}

The SER hyperparameter $\sigma_0^2$ can be fixed, or it can be
estimated by maximizing the approximate (partial)
likelihood, which is given by
\begin{align}
\hat{\ell}_{\mathrm{SER}}(\sigma_0^2) =
\sum_{j=1}^p \pi_j \textstyle \int \hat{\ell}(b; \x_{\cdot j}, {\bm c}) \,
p(b \mid \sigma_0^2) \, db.
\end{align}
A simple expectation maximization (EM) algorithm \citep{em} can be
used to maximize $\hat{\ell}_{\mathrm{SER}}$. This EM algorithm cycles
between computing the posterior quantities \eqref{eq:coxph-ser} at a
given $\sigma^2$ (this is the E-step), then updating $\sigma_0^2$ by
\begin{align}
\sigma_0^2 \leftarrow \sum_{j=1}^p\alpha_j(\mu_{1j}^2 + \sigma_{1j}^2)
\label{eq:sigma0-em-update}
\end{align}
(this is the M-step). See the Appendix
for a derivation of this EM algorithm.


\subsection{CoxPH-SuSiE}

A key feature of the SuSiE model from \cite{wang2020simple} is that,
given estimates of $\b_1, \dots, \b_{L-1}$, estimating $\b_L$
corresponds to fitting a Gaussian SER model with offset ${\bm c} =
\sum_{l=1}^{L-1} \X \b_l$. As \cite{wang2020simple} notes, this
suggests an {\em iterative approach} to fitting a SuSiE model: the
idea is to cycle through the $L$ individual SER models, updating the
offset and fitting a Gaussian SER model. They called this iterative
model fitting procedure ``Iterative Bayesian Stepwise Selection''
(IBSS). Here we generalize these ideas to fit the CoxPH-SuSiE model,
resulting in an iterative algorithm which in each iteration of the
algorithm fits a CoxPH-SER model. We call the resulting algorithm
``generalized Iterative Bayesian Stepwise Selection'' (gIBSS); it is
summarized in Algorithm \ref{algo:gibss}.

\SetKwInput{KwRequire}{Require}
\SetKwInput{KwReturn}{Return}

\begin{algorithm}[t]
  \normalsize
  
\KwRequire{Data inputs $\X$ ($n \times p$ matrix), $\y$ ($p$-vector of
  observed times), and $\bdel$ ($p$-vector, 0 = censored, 1 =
  uncensored).}

\KwRequire{$L$, the number of single effects.}

\KwRequire{Initial estimates of the prior variances, $\sigma_{0l}^2$,
$l = 1, \ldots, L$.}

\KwRequire{A function $\text{\sc CoxPH-SER}(\X, \y, \bdel, {\bm c};
  \sigma_0^2) \rightarrow (\balpha, \bmu_1, \bsig_1^2)$ that computes
  an (approximate) posterior distribution of $(b, {\bm\gamma})$ for an
  CoxPH-SER model with prior variance $\sigma_0^2$ given data $\X, \y,
  \bdel, {\bm c}$.}

Initialize the $p$-vectors of posterior mean coefficients, $\bar{\bm b}_l$,
$l = 1, \ldots, L$.

Initialize the $n$-vector of offsets, ${\bm c} = {\bm 0}$.

\Repeat{\rm convergence criterion is met or the maximum number of iterations reached}{
  \For{\rm $l$ in $1, \ldots, L$}{

  Remove the $l$th single effect from the offsets,
  ${\bm c}_l \leftarrow {\bm c} - \X \bbar_l$.

  $(\balpha_l, \bmu_l, \bsig_l^2) \leftarrow
  \text{\sc CoxPH-SER}(\X, \y, \bdel, {\bm c}_l; \sigma_{0l}^2)$ 
  
  Update $\sigma_{0l}^2$ using \eqref{eq:sigma0-em-update},
  $\sigma_{0l}^2 \leftarrow \sum_{j=1}^p
  \alpha_{jl}(\mu_{jl}^2 + \sigma_{jl}^2)$.
  
  Update the posterior mean coefficients, $\bbar_l = \balpha_l \circ \bmu_l$.

  Update the offsets, ${\bm c} \leftarrow {\bm c}_l + \X \bbar_l$.
  }
}

\KwReturn{CoxPH-SuSiE posteriors
  $\balpha_1, \ldots, \balpha_L, \bmu_1, \ldots, \bmu_L,
  \bsig_1^2, \ldots, \bsig_L^2$}
\, \\

\caption{Generalized Iterative Bayesian Stepwise Selection for CoxPH-SuSiE.}
\label{algo:gibss}
\end{algorithm}

gIBSS returns approximate posterior distributions for the single
effect vectors $\b_l$. We use these approximate posterior
distributions to compute, for each $l$, a point estimate of $\b_l$ (we
typically use the posterior mean, $\bar{\b}_l$) and a Credible Set
(CS). A CS is a set of variables that has a high probability of
containing the variable with the non-zero regression
coefficient. 
\begin{definition}[\citealt{wang2020simple}]
\label{def:cs}
A level-$\rho$ Credible Set is a subset of variables that has
probability $\ge \rho$ of containing at least one effect variable
(i.e., a variable with non-zero regression coefficient).
\end{definition}
Given $\balpha_l$, it is straightforward to construct the CS. First,
sort the variables in descending order by $\alpha_{lj}$. Then, add
variables to the CS until their cumulative probability exceeds
$\rho$. In practice, we may further prune the CSs based on their
``purity'', which is defined as the smallest absolute correlation
among all pairs of variables within the CS. The reason for doing this
is that a CS may lack inferential value because it contains many
uncorrelated variables.

While the gIBSS algorithm follows the same logic as the IBSS algorithm
of \cite{wang2020simple}, there is an important difference: IBSS can
be rigorously justified as optimizing a variational approximation to
the posterior distribution of $\b$ under the (Gaussian) SuSiE model
\citep{wang2020simple}, but we are not able to provide a similar
result for gIBSS for fitting CoxPH-SuSiE models. As a result, we are
not able to prove that gIBSS is optimizing a specific objective
function, or that the gIBSS iterations are guaranteed to converge
(although we find that they generally do in practice). We therefore
view gIBSS as a heuristic generalization of IBSS, and rely on
numerical experiments to validate the approximate inferences and
demonstrate its good performance in practice.

\subsection*{Computational complexity}
\label{sec:complexity}

The predominant computation in gIBSS is the repeated fitting of
single-variable CoxPH regression models to obtain $\hat{b}, s^2$
(Section \ref{sec:bayes_simple}). Within a single gIBSS iteration,
these computations are performed once for each of the $L$ single
effects and once for each of the $p$ covariates. Since these
computations scale linearly with $n$, the per-iteration computational
complexity of gIBSS is $O(npL)$. Although this is the same as the
computational complexity of (Gaussian) SuSiE, unlike Gaussian SuSiE the
CoxPH simple single-variable regressions do not admit closed-form
solutions, and so CoxPH-SuSiE is considerably slower than Gaussian
SuSiE.

Note that fitting the univariate regressions can easily be
performed in parallel, so CoxPH-SuSiE can take advantage of multicore
architectures to greatly speed up the gIBSS algorithm. We exploited
this property when we applied CoxPH-SuSiE to the large UK Biobank data
sets in Section~\ref{sec:coxph_susie_real_data}.

\subsection*{Extension for additional covariates}

In a genetic association or fine-mapping analysis, it is often desired
to include covariates such as sex and ``genetic principal components''
to reduce confounding due to population structure \citep{price2006pca,
  bycroft2018uk}. In principle, CoxPH-SuSiE is naturally extended to
allow for $m$ additional covariates by including additional terms in
the single-variable CoxPH model \eqref{eq:coxph}:
\begin{equation}
\begin{aligned}
\lambda_i(t) &= \lambda_0(t)\exp(\bm{w}^{\intercal}\bm{z}_i + bx_i + c_i) \\
b&\sim N(0, \sigma_0^2) \\
\bm{w} &\sim N(0, \bm{\Sigma}_w),
\end{aligned}
\label{eq:coxph-extended}
\end{equation}
Here, $\bm{z}_i$ denotes the additional covariates for sample $i$ (a
vector of length $m$), the corresponding coefficients are denoted as
$\bm{w}$, and $\bm{\Sigma}_w$ denotes the prior covariance of
$\bm{w}$. However, including these additional covariates in the CoxPH
model means that the 1-d integrals in the posterior computations for
the single-variable SER would become $(m + 1)$-dimensional
integrals. This would greatly increase the computational expense of
applying CoxPH-SuSiE to large data sets (such as the UK Biobank data
we analyze in Section \ref{sec:coxph_susie_real_data}). Therefore, we
implemented the following practical approach based on an approximation
similar to the approximation used in CoxPH-SuSiE: first we fit a CoxPH
regression model that {\em only} included the additional
covariates---that is, \eqref{eq:coxph-extended} with $b = 0$ and $c_i
= 0$, $i = 1, \ldots, n$---then we initialized the offsets $c_i$ in
the gIBSS algorithm to $c_i = \hat{\bm{w}}^{\intercal}\bm{z}_i$, where
$\hat{\bm{w}}$ was the vector estimated coefficients. This approach imposed
only a one-time cost of fitting a CoxPH regression model with $m$
covariates. We used this approach in the CoxPH-SuSiE fine-mapping
analyses of asthma, where we included additional covariates for sex
and genetic PCs (Section \ref{sec:coxph_susie_real_data}).

\subsection{Alternative approaches}
\label{sec:alternative}

While developing the approach described above, we also explored
alternative approaches to fitting CoxPH-SuSIE models. In one approach,
we approximated the Bayes factors using the Asymptotic Bayes Factor
(ABF) approximation from \cite{wakefield2009bayes} instead of the
Laplace BF \eqref{eq:BF_lap}. Both approximations use the
maximum-likelihood estimate $\hat{b}$ and its variance $s^2$, but they
use these quantities in different ways. In our experiments (Section
\ref{sec:BFcomp}), we found that the Laplace BF was more accurate than
the ABF, so we used the Laplace BF in our CoxPH-SuSiE method.

Another approach we considered built on the ``SuSiE-RSS'' method from
\cite{zou2022fine}. SuSiE-RSS fits the same Gaussian SuSiE model as
\cite{wang2020simple}, but the computations are different because
SuSiE-RSS works with summary statistics instead of individual-level
data; specifically, SuSiE-RSS accepts the least-squares estimate of
the coefficient, $\hat{b}$, and its variance, $s^2$, for each of the
$p$ covariates, and the $p \times p$ sample correlation matrix. A
simple idea then is to provide SuSiE-RSS with the $\hat{b}, s^2$
obtained from the $p$ CoxPH regression models. This approach is
potentially attractive because the single-variable CoxPH
regression models need only to be fit once, rather than once per
iteration and per single effect in gIBSS. However, in our experiments
(see below) this approach performed considerably worse than
gIBSS. Note that in the $L = 1$ case, applying SuSiE-RSS in this way
corresponds to using the ABF instead of the Laplace BF, and as we
noted above the ABF was less accurate than the Laplace BF, providing
an additional point against this SuSiE-RSS-based approach.

\section{Comparison of different approaches to computing the Bayes factors}
\label{sec:BFcomp}

In our first set of experiments, we empirically assessed the accuracy
of the Laplace Bayes Factor \eqref{eq:BF_lap} and compared it to the
Asymptotic Bayes Factor \eqref{eq:abf}. To provide a ``gold standard''
to compare to, we computed accurate estimates of the Bayes factors
\eqref{eq:BFp} using numerical quadrature methods. [We used
  Gauss-Hermite quadrature implemented by the {\tt gauss.quad.prob}
  function from the statmod R package \citep{statmod}, with 32 nodes;
  for details, see the Appendix.] For all the Bayes factors, we fixed the
prior variance $\sigma_0^2$ to 1, and we set the offset to zero.

Since we were ultimately interested in applying these methods to data
from UK Biobank with very large sample size ($n \approx
\mbox{500,000}$), we simulated genetic data sets with $n =
\mbox{500,000}$ to model this setting. For each simulation, we
generated censored TTE data from a ``single-SNP'' CoxPH model: we a
simulated genetic variant---specifically, a single nucleotide
polymorphism (SNP)---as $x_i \sim \mathrm{Binom}(2,f)$, where $f$ was
the minor allele frequency (MAF), then we simulated the TTE phenotype
from the single-SNP CoxPH model with regression coefficient $b =
0.1$. We simulated data sets with different combinations of censoring
rates and MAFs; we simulated 50 data sets for each
combination. Additional details on how these censored TTE data were
simulated are given in the Appendix.


\begin{figure}[t]
\centering
\includegraphics[width=0.925\textwidth]{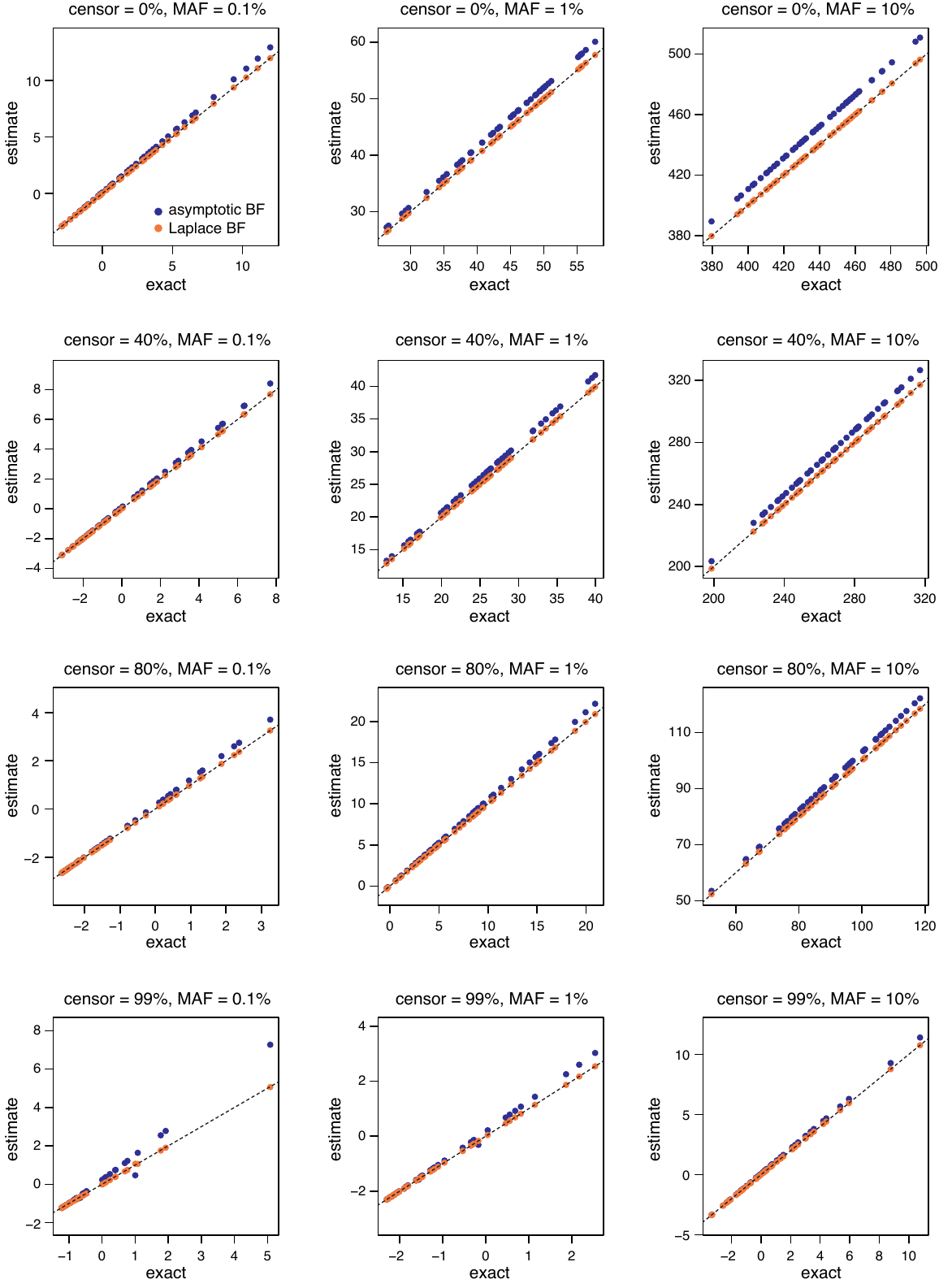}
\caption{\rm Comparison of (log base-10) Bayes factor estimates in
  single-SNP simulations with varying censoring rates and SNP minor
  allele frequencies (MAFs). Each plot shows results from 50
  simulations. ``Exact'' is the numerical quadrature estimate of the
  Bayes factor (which is not necessarily exact, but should be very
  accurate).}
  %
  %
\label{fig:bf_comparison}
\end{figure}

In these simulations, the Laplace BF was very accurate for all
censoring rates and MAFs
(Figure~\ref{fig:bf_comparison}). Importantly, the Laplace BF achieved
good accuracy at a much lower computational cost than the
computationally intensive numerical quadrature method.
%
%
By contrast, the ABF was less accurate in simulations, with a tendency
to overestimate the BF. Since the effort involved in computing the
Laplace BF and ABF is very similar,
%
%
the Laplace BF is clearly preferable, and so this is what we
used to implement CoxPH-SuSiE.

\section{Comparison of BVSR methods for fine-mapping
  TTE phenotypes}
\label{sec:simulations}


We simulated fine-mapping data sets to assess the performance of
CoxPH-SuSiE for the task of fine-mapping TTE phenotypes. In these
simulations, we also compared CoxPH-SuSiE against three existing BVSR
methods for censored TTE outcomes for which software was available:
R2BGLiMS \citep{newcombe2017weibull}, BVSNLP
\citep{nikooienejad2020bayesian} and survival.svb
\citep{komodromos2022variational}. These three methods were
implemented as R packages \citep{R} and we refer to the methods by the
names of their R packages. We also assessed the SuSiE-RSS-based
approach which was described in Section \ref{sec:alternative}.

To mimic realistic fine-mapping settings, we simulated censored TTE
phenotypes using real genotype data from the Genotype-Tissue
Expression (GTEx) project \citep{gtex2015genotype, gtex2017} and UK
Biobank \citep{sudlow2015uk, bycroft2018uk}.  The genotype data from
GTEx and UK Biobank are both well suited to illustrate fine-mapping
due to the high density of available genetic variants. The UK Biobank
data in particular are of a much larger sample size and therefore are
helpful for assessing the ability of the methods to cope with
large-scale data. Both data sets feature genetic variants with
complex patterns of correlation and many very strong correlations.
Therefore, we did not expect any of the methods to achieve 100\%
accuracy in identifying the causal variants, even with the larger
sample sizes.

\begin{table}[t]
\centering
\sf\footnotesize
\begin{tabular}{ccccc}
\toprule 
& & covariates & effect & average \\
genotypes & sample size & (SNPs) & variance & region size \\ 
\midrule
GTEx & 574 & 1,000 & 1 & 280 kb \\
UK Biobank & 50,000 & 1,000 & 0.1 & 300 kb \\
\bottomrule
\end{tabular}
\vspace{1ex}
\caption{\rm Summary of the settings for the fine-mapping simulations.
\label{tab:sim_param}}
\end{table}

We used the GTEx and UK Biobank genotype data to generate two sets of
simulations, which are summarized in Table \ref{tab:sim_param}.  For
each of the GTEx simulations, we randomly selected a fine-mapping
region containing 1,000 SNPs between base-pair positions 61,597,515
and 63,597,178 on chromosome 1 (Genome Reference Consortium human
genome assembly 38). On average, the region containing 1,000 SNPs
spanned 280 kb.  For each of the UK Biobank simulations, we randomly
subsampled 50,000 genotype samples, then we selected a region
containing 1,000 SNPs between base-pair positions 18,510,134 and
19,065,757 on chromosome 3 (Genome Reference Consortium human genome
assembly 37, hg19). On average, the 1,000-SNP region spanned
300 kb.

%
%

For each simulation, we randomly selected a small number of SNPs to be
the SNP affecting the TTE outcome, then we simulated censored TTE data
following the procedure described in the Appendix. In both sets of
simulations, we simulated data sets with different numbers of causal
variables (ranging from 1 to 3, as well as the setting in which none
of the causal variables affected the phenotype) and different
censoring levels. In total, we simulated 400 fine-mapping data sets
using the GTEx genotypes and 240 data sets using the UK Biobank data
sets. See the Appendix for more details.
%
%
%
%
%
%


Each of the methods has several tuning parameters that may affect the
performance and running time of the method; investigating the impact
of the tuning parameters is beyond the scope of this experiment, so
when possible we followed the guidance given in the publications and
in the software documentation. We also adjusted some settings so that
the running time of the methods was similar. (Except for SuSiE-RSS,
which is much faster than all the other methods for data sets with
large sample size because it uses precomputed summary statistics.)
For BVSNLP, we called the {\tt bvs()} function with {\tt prep = FALSE}
to skip the internal data preprocessing step, and we used a
Beta-binomial prior for the model space. For R2BGLiMS, we called the
{\tt R2BGLiMS()} function with the Weibull modeling option and a
Beta-binomial prior for the model space.  For survival.svb, we called
the {\tt svb.fit()} function with the maximum number of iterations set
to 1,000 for the GTEx data sets and lowered it to 100 for the UK
Biobank data sets to reduce running time.  For CoxPH-SuSiE, we used
the {\tt ibss\_from\_ser()} function from the logisticsusie R package
with the number of single effects, $L$, set to 5 and the prior
variance $\sigma_0^2$ was initialized to 1. (The logisticsusie R
package is available at
\url{https://github.com/karltayeb/logisticsusie/} and in the
Zenodo repository; \citealt{zenodo}.) We set the
maximum number of iterations to 100 for the GTEx data sets and 10 for
the UK Biobank data sets. For SuSiE-RSS, we ran the {\tt susie\_rss()}
function from the susieR package \citep{wang2020simple, zou2022fine}
with {\tt L = 5}, {\tt var\_y = 1} and {\tt max\_iter = 100}. For both
CoxPH-SuSiE and SuSiE-RSS, the required summary statistics --- that
is, the coefficient estimates and their corresponding variances ---
were computed using the {\tt coxph()} function from the survival R
package \citep{survival-book}.\footnote{The
likelihood ratio $\ell(\hat b; {\bm x}, {\bm c})/\ell(0; {\bm x}, {\bm
  c})$ is also needed to compute the Laplace BFs \eqref{eq:BF_lap} in
CoxPH-SuSiE. The survival package provides this quantity for the CoxPH
model.} Other tuning parameters not mentioned were kept at their
defaults for all methods.  The code used to run these experiments can
be found in a git repository on GitHub,
\url{https://github.com/yunqiyang0215/survival-susie/} (see also
\citealt{zenodo}). Other details, including versions of the
software used and the computing setup, are given in the Appendix.

As mentioned in the introduction, the two key statistical
quantities for fine-mapping are PIPs and CSs. However, since most of
the methods do not yield CSs, we focused our comparisons on
PIPs. (CoxPH-SuSiE and SuSiE-RSS provide both PIPs and CSs, and we
compare the CSs from these two methods below.)

\begin{figure}[t]
\centering\includegraphics[width=1\linewidth]{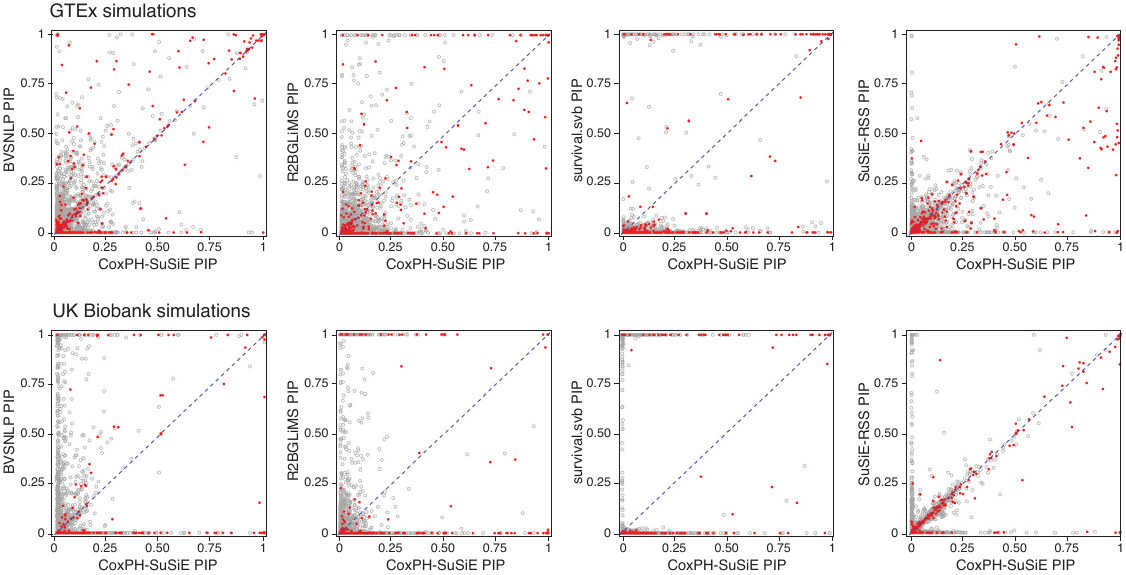}
\caption{\rm CoxPH-SuSiE PIPs vs. PIPs from other methods.
  Each point is a variable (SNP); true causal variables
  (causal SNPs) are shown as solid red circles, and non-causal
  variables are shown as open gray circles. Each plot in the top row
  contains 400,000 points (1,000 SNPs $\times$ 400 data sets), and
  each plot in the bottom row contains 240,000 points (1,000 SNPs
  $\times$ 240 data sets). See Supplementary Figures
  \ref{fig:pip_gtex} and \ref{fig:pip_ukb} 
  for a breakdown of these results by censoring level.}
\label{fig:pip_combined}
\end{figure}

\begin{figure}[t]
\centering
\includegraphics[width=\textwidth]{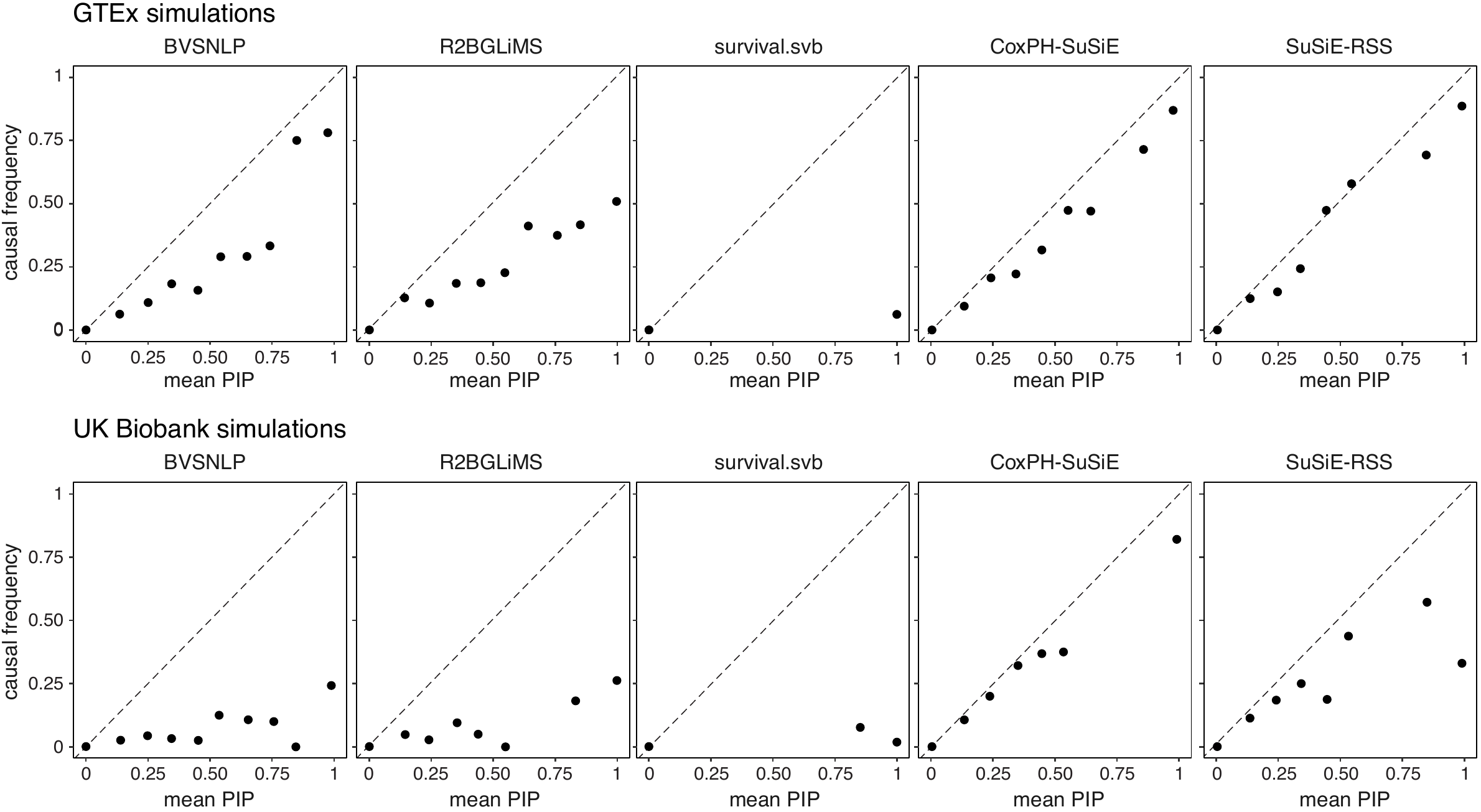}
\caption{\rm Assessment of PIP calibration. SNPs from all simulations
  (400 GTEx simulations and 240 UK Biobank simulations) were grouped into
  bins according to their PIP values (10 equally spaced bins from 0 to
  1). The plots then show the average PIP from each bin ($x$-axis)
  against the proportion of SNPs in that bin that are causal ($y$-axis).
  Bins with fewer than 10 observations were removed from the plots. A
  well-calibrated method should produce a plot with points near the
  diagonal.}
    \label{fig:pip_calibration_combined}
\end{figure}

\begin{figure}[t]
\centering
\includegraphics[width=0.8\textwidth]{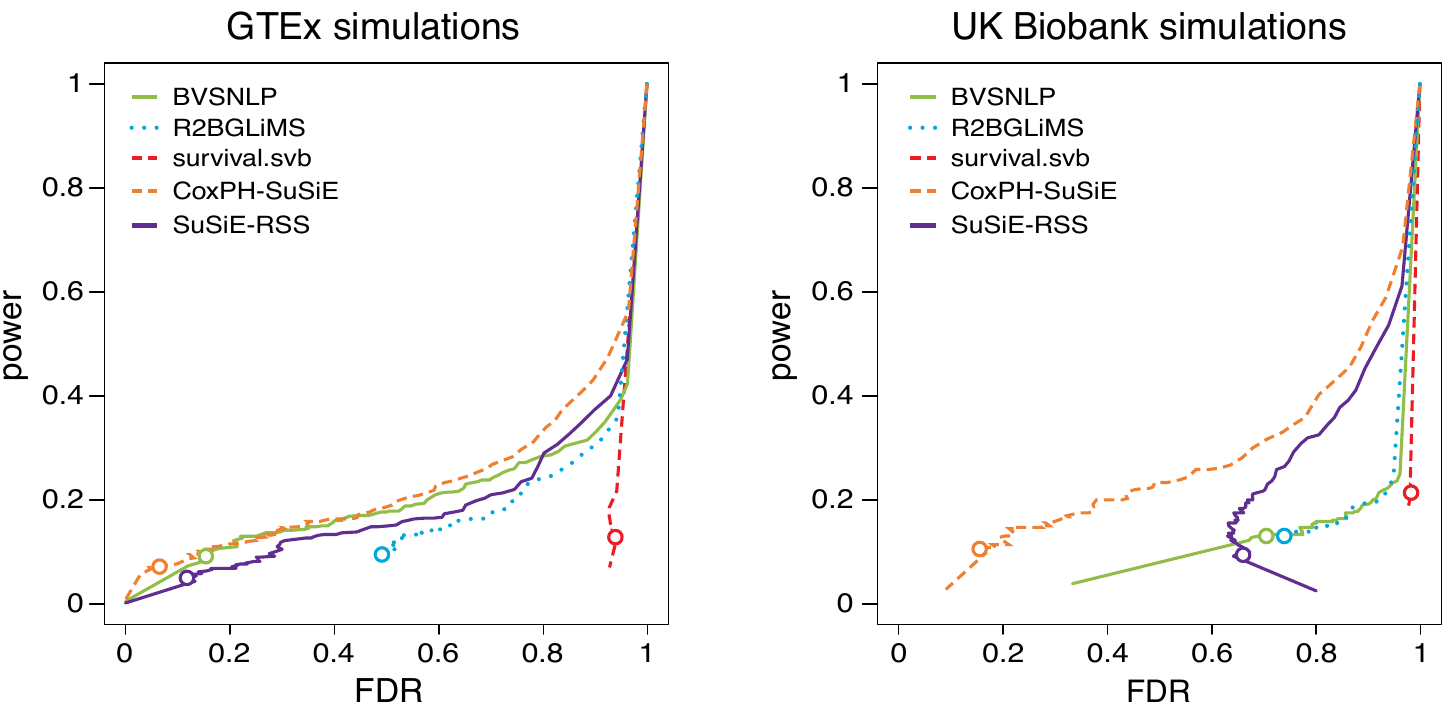}
\caption{\rm Discovery of causal SNPs using posterior inclusion
  probabilities (PIPs). Each curve shows power vs. FDR in identifying
  causal SNPs. FDR and power are defined as $\text{FDR} \colonequals
  \text{FP}/(\text{TP} + \text{FP})$ and $\text{power} \colonequals
  \text{TP}/(\text{TP} + \text{FN})$, where FP, TP, FN and TN denote
  the number of false positives, true positives, false negatives and
  true negatives, respectively. FDR and power were calculated from
  400 GTEx simulations and 240 UK Biobank simulations 
  as the PIP threshold was varied from 0 to
  1.  Note that these plots are the same as precision-recall plots
  after flipping the $x$-axis because $\text{precision}
  %
  %
  = 1 - \text{FDR}$ and recall = power.)  Open circles are drawn at a
  PIP threshold of 0.95. See Supplementary Figures \ref{fig:fdr_gtex}
  and \ref{fig:fdr_ukb} for the power vs. FDR results stratified
  by censoring level.}
\label{fig:fdr_combined}
\end{figure}

First, we found that the different methods, although all taking
conceptually similar approaches to the problem, produced strikingly
different PIPs (Figure~\ref{fig:pip_combined}). So we examined these
differences more closely. One hope is that the PIPs are {\em well
  calibrated}; e.g., that among SNPs assigned a PIP near 0.95,
approximately 95\% should be causal. Among the methods compared, the
CoxPH-SuSiE PIPs were best calibrated in both simulation settings,
followed by SuSiE-RSS (Figure~\ref{fig:pip_calibration_combined}).
%
%
Correspondingly, most methods overstated the PIPs, particularly in the UK
Biobank simulations: among SNPs with PIP 
95\% or greater, at most 35\% actually were causal, 
except for CoxPH-SuSiE, which identified causal SNPs 84\% of the time
(which is much better than the other methods, but still leaves 
some room for improvement). 
Among all the methods compared, survival.svb was least well
calibrated: almost all its PIPs were concentrated near 0 or 1, 
and the PIPs above 95\% were almost always false positives
(98\% of the time).

\begin{table}[t]
\centering
\sf\footnotesize
\begin{tabular}{cr@{\;}lr@{\;}l}
\toprule 
method & \multicolumn{2}{c}{GTEx} & \multicolumn{2}{c}{UK Biobank} \\ 
\midrule
BVSNLP & 58 & (15.7--2,042) & 4,499 & (490--79,601) \\
R2BGLiMS & 190 & (167--494) & 10,463 & (8,870--14,118) \\
survival.svb & 139 & (6--844) & 2,496 & (118--15,353) \\
CoxPH-SuSiE & 331 & (80--1,014) & 7,275 & (4,472--14,416) \\
SuSiE-RSS & 6.4 & (6.0--7.6) & 218 & (199--296) \\
\bottomrule
\end{tabular}
\vspace{1ex}
\caption{\rm Running times in seconds of the different methods in the
  fine-mapping simulations. The first number is the average across all
  simulations; the numbers in parentheses give the full range across the
  simulations.
\label{tab:sim_runtimes}}
\end{table}

We also compared how effectively the PIPs from different methods
identified the causal SNPs by comparing the power and FDR of each
method at different PIP thresholds (Figure~\ref{fig:fdr_combined}). In
the GTEx simulations, CoxPH-SuSiE PIPs almost always had highest power
at each FDR threshold, followed closely by BVSNLP and SuSiE-RSS. In
the UK Biobank simulations, the CoxPH-SuSiE PIPs far outperformed the
PIPs from the other methods. The survival.svb PIPs exhibited
particularly poor performance in both simulation settings. As
expected, power for all methods went down as censoring increases, but
CoxPH-SuSiE maintained the best or among the best performance in all
censoring settings (Supplementary Figures~\ref{fig:fdr_gtex} and
\ref{fig:fdr_ukb}).

While CoxPH-SuSiE does require considerable computational effort to
run on large data sets, its running times were nonetheless quite
comparable to the other methods (Table \ref{tab:sim_runtimes}). The
one exception was SuSiE-RSS, which uses only summary statistics, and
therefore is comparatively fast for data sets with large sample sizes.

%
%
%


\begin{figure}
\centering
\includegraphics[width=\textwidth]{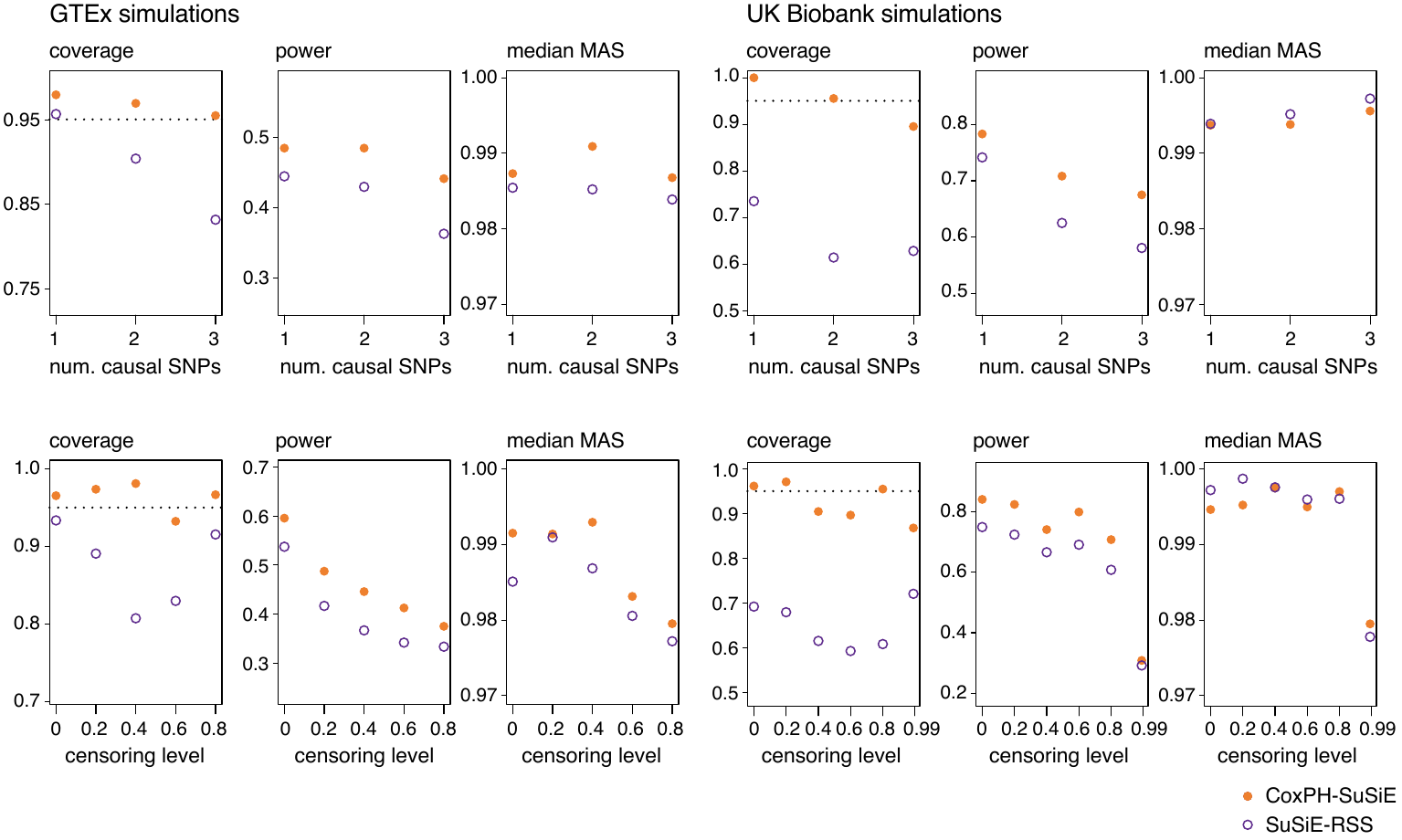}
\caption{\rm Assessment of 95\% credible sets (CSs) from CoxPH-SuSiE
  and SuSiE-RSS. The performance metrics are: {\em coverage}, the
  proportion of CSs that contain a true causal SNP; {\em power}, the
  proportion of true causal SNPs included in a CS; and {\em median
    MAS}, in which {\em MAS} is ``mean absolute correlation'', the
  mean of the correlations (Pearson's {\em r}) among all pairs of SNPs
  within the given CS. The results are stratified by number of causal
  SNPs in the top row and by censoring level in the bottom row. The
  target coverage (95\%) is shown as a dotted horizontal
  line. Following \cite{wang2020simple, zou2022fine}, all CSs returned
  by CoxPH-SuSiE and SuSiE-RSS with ``purity'' (the minimum absolute
  correlation among all pairs of SNPs) less than 0.5 were not
  considered.}
  %
  %
\label{fig:cs}
\end{figure}

Among the methods compared, only CoxPH-SuSiE and SuSiE-RSS provide
CSs. Each CS is meant to capture at least one causal variable (with
high probability). A key performance metric is {\em coverage}, the
proportion of CSs that contain at least one true causal SNP. We found
that, across the different censoring levels and numbers of causal
SNPs, the CoxPH-SuSiE CSs achieved, or came very close to, the target
coverage of 95\%, whereas the SuSiE-RSS CSs performed less well, with
coverage dropping as low as 0.6 (Figure~\ref{fig:cs}). The good
coverage of CoxPH-SuSiE CSs did not come at the cost of the other
performance metrics: in most scenarios CoxPH-SuSiE outperformed SuSiE
RSS in both power and median MAS (Figure~\ref{fig:cs}).



\section{Application to fine-mapping asthma loci in UK Biobank}
\label{sec:coxph_susie_real_data}

To demonstrate our method on real data, we analyzed 
age of diagnosis data for asthma 
%
%
in UK Biobank samples
\citep{sudlow2015uk}. The UK Biobank is a large, population-based
prospective study, with detailed phenotype and genotype data from over
500,000 participants in the United Kingdom (ages were between 40 and
69 at time of recruitment). The UK Biobank imputed genotypes feature a
high density of available SNPs \citep{bycroft2018uk}, so they are well
suited for fine-mapping. Several previous studies have performed
association analyses for asthma using UK Biobank data
\citep{clay2022asthma, ferreira2019genetic, han2020asthma,
  pividori2019shared, valette2021asthma, vincente2017asthma,
  zhu2018asthma}. This includes \cite{bi2020fast}, who used a CoxPH
model to perform their association analysis.
%
%
But, to our knowledge, only \cite{zhu2018asthma, clay2022asthma,
  zhong2025integration} took the step of fine-mapping the asthma
loci. (Note that \citealt{clay2022asthma} focused on fine-mapping the
HLA region.) None of these fine-mapping analyses exploited the
available TTE information (that is, the age at which asthma was
diagnosed).

%


\begin{figure}[t]
\centering
\includegraphics[width=\textwidth]{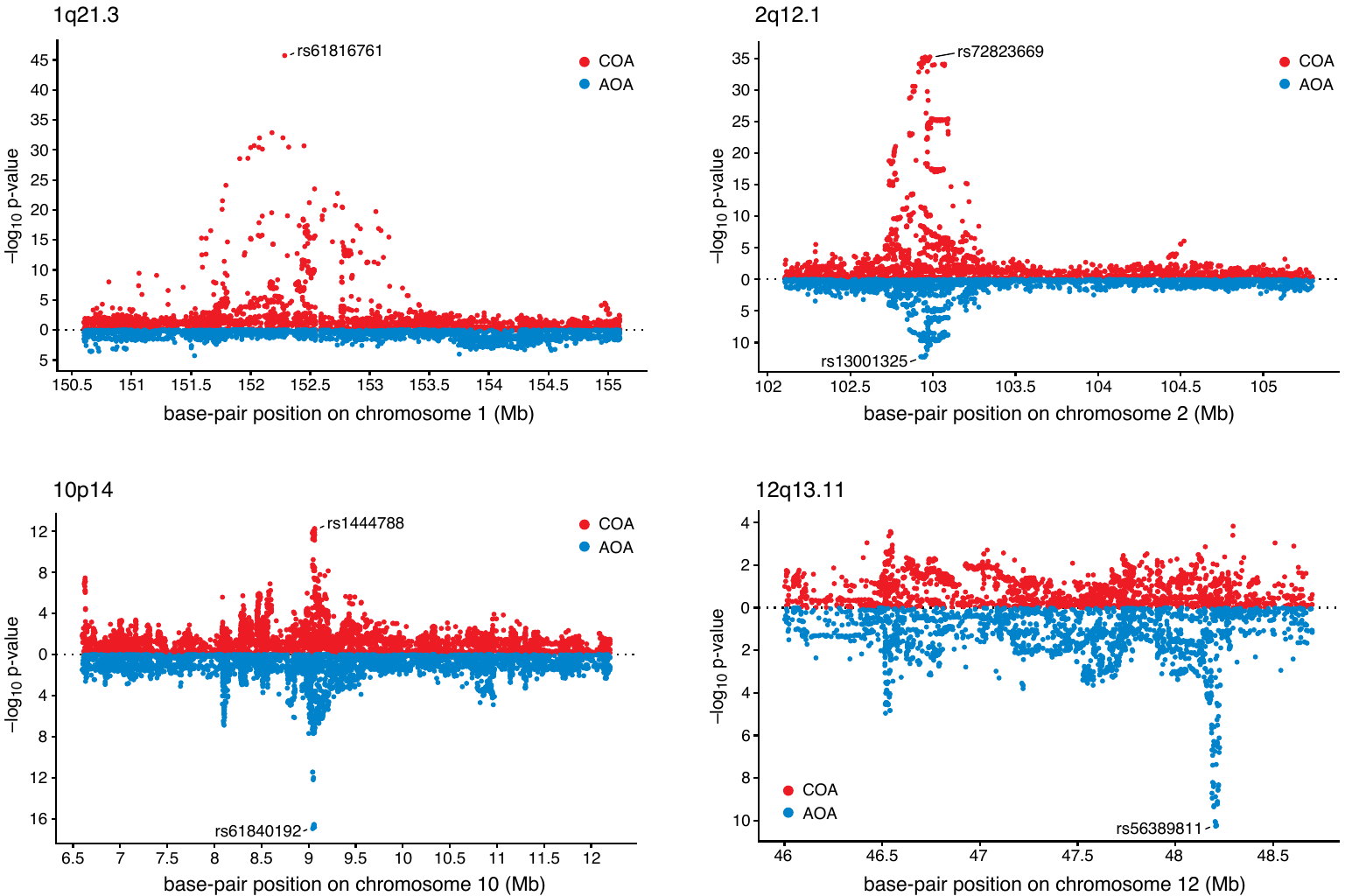}
\caption{\rm CoxPH-based association analyses of the UK Biobank TTE
  asthma data for four of the asthma loci identified in
  \cite{pividori2019shared}. Each point depicts a SNP. At each locus,
  the top association for COA and/or AOA is highlighted. The {\em
    p-}values were obtained from SPACox \citep{bi2020fast},
  specifically the {\tt p.value.spa} column of the SPACox output. Only
  SNPs with MAF greater than 1\% were included in the association
  analyses.
  %
  %
  See Table \ref{tab:asthma_data} and the Appendix for details on the
  data used in these association analyses.}
\label{fig:aoa_coa_gwas}
\end{figure}

\begin{table}[t]
%
%
%
\centering
\sf\footnotesize
\begin{tabular}{lrr}
trait & observed & censored \\
\midrule
COA & 8,024 & 260,805 \\ 
AOA & 18,569 & 242,399 \\
AA & 31,860 & 241,683 \\ 
\bottomrule
\end{tabular}
\vspace{1ex}
\caption{\rm Summary of the TTE data used in the asthma association
  analyses and asthma fine-mapping analyses.
\label{tab:asthma_data}}
\end{table}


\begin{table}[t]
\centering
\sf\footnotesize
%
%
\begin{tabular}{lclcrr}
\toprule
locus & traits & top SNP  & region (bp) & size (kb) & SNPs \\
\midrule
1q21.3 & COA & rs61816761 & 152,037,453--152,535,675 & 498.22 & 1,333 \\
2q12.1 & COA, AOA & rs72823641 & 102,686,430--103,182,687 & 496.26 & 1,821 \\
2q22.3 & AOA & rs7571606 & 145,873,121--146,372,665 & 499.54 & 1,190 \\
10p14 & COA, AOA & rs2197415 & 8,813,629--9,312,765 & 499.14 & 1,651 \\
11q13.5 & COA, AOA & rs11236797 & 76,050,271--76,549,513 & 499.24 & 1,628 \\
12q13.11 & AOA* & rs56389811 & 47,955,425--48,455,255 & 499.83 & 1,595 \\
15q22.2 & COA, AOA & rs11071559 & 60,820,809--61,319,821 & 499.01 & 1,558 \\
17q12 & COA & rs4795399 & 37,812,435--38,311,433 & 499.00 & 1,212 \\
\bottomrule
\end{tabular}
\vspace{1ex}
\caption{\rm Asthma loci from \cite{pividori2019shared} fine-mapped
  using CoxPH-SuSiE. The ``traits'' column gives the disease
  phenotypes (COA and/or AOA) that have significant associations in
  the region. The ``top SNP'' is the SNP with the strongest
  association (smallest SPACox association test {\em p-}value) in that
  region for either trait. The chromosomal base-pair positions are
  based on human genome assembly 19 (Genome Reference Consortium Human
  Build 37, February 2009) *The 12q13.11 locus was classified as both
  COA and AOA in \cite{pividori2019shared} based on additional
  analyses, but showed significant association only in AOA so we
  treated it as AOA only.
  %
  %
\label{tab:asthma_loci}}
\end{table}

Here, we fine-mapped asthma loci by applying CoxPH-SuSiE to the asthma
age-of-diagnosis data from UK Biobank. (Specifically, these were
self-reported doctor diagnoses; see the Appendix for details on how
event times and censoring status were defined.) We selected 8 loci
(regions) for fine-mapping from the associations reported in
\cite{pividori2019shared} (Table \ref{tab:asthma_loci}).
\cite{pividori2019shared} distinguished between genetic associations
for childhood-onset asthma (COA) and adult-onset asthma (AOA), and we
selected loci that showed a variety of association patterns: some with
associations with both AOA and COA, and others with associations in
only one of these two. We performed our own association analyses using
a CoxPH model to confirm these patterns of association also hold in a
TTE analysis (Figure~\ref{fig:aoa_coa_gwas}).

If a causal SNP is specific to COA, then including AOA events in the
fine-mapping analysis could introduce noise rather than enhance
power. Conversely, if a SNP affects both AOA and COA risk, analyzing
all asthma events together may be more powerful.  We therefore
fine-mapped each region using one of three analysis strategies: using
childhood onset events for regions showing COA associations only;
using adult onset events (ignoring childhood onset cases) for those
regions showing AOA associations onlyl and using all asthma (AA) onset
events for those regions showing both COA and AOA associations. We
used the definitions of COA and AOA from \cite{pividori2019shared}. At
each locus, we included all SNPs with MAFs of 1\% or more that were
within 250 kb of the top association. We ran CoxPH-SuSiE with $L = 10$
single effects and filtered out CSs with purity less than 0.5. To cope
with the large scale of the UK Biobank data, we stopped the gIBSS
algorithm after 10 iterations (if the stopping criterion had not
already been met). See the Appendix for details on the steps taken to
prepare the UK Biobank phenotype and genotype data for the CoxPH-SuSiE
analyses. All the code used to generate the results presented here can
be accessed at
\url{https://github.com/yunqiyang0215/survival-data-analysis}. This
code is also included in the Zenodo repository \citep{zenodo}.

\begin{figure}[t]
\centering
\includegraphics[width=\linewidth]{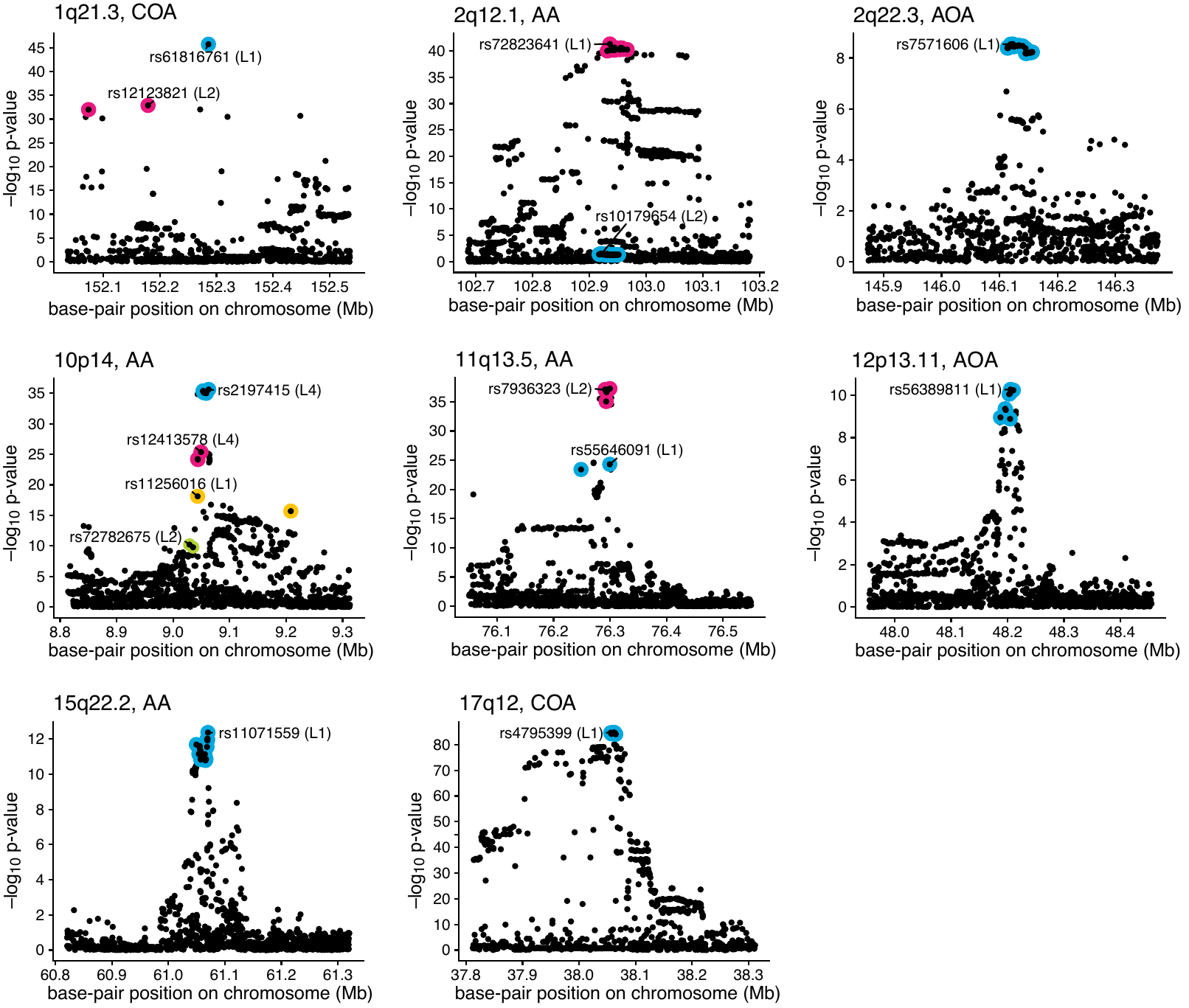}
\caption{\rm CoxPH-SuSiE asthma fine-mapping results.
  %
  %
  Each point depicts a single SNP. The {\em p-}values were computed by
  SPACox \citep{bi2020fast}. Membership of SNPs to CSs is indicated by
  different colors. The ``sentinel SNP'' (i.e.,~the SNP with the
  highest PIP) in each CS is labeled.}
\label{fig:susie1}
\end{figure}

%
%

\begin{figure}[t]
\centering
\includegraphics[width=0.35\textwidth]{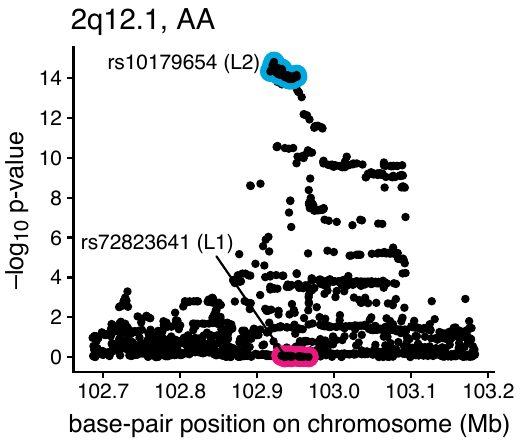}
\caption{\rm CoxPH-based association analysis of the 2q12.1 asthma
  locus conditioned on the genotype of a candidate asthma SNP at the
  same locus. The association tests are conditioned on the genotype of
  SNP rs72823641, which is the sentinel SNP of the first CS (see
  Figure~\ref{fig:susie1} and Table \ref{tab:susie}). The {\em p-}values
  were computed using SPACox, with the genotype of SNP rs72823641
  included as an additional covariate. Each point depicts a SNP. The
  two CSs returned by CoxPH-SuSiE (Figure~\ref{fig:susie1}) are shown by
  the red and blue circles.}
\label{fig:cond_pval}
\end{figure}

\begin{table}[t]
\centering
\sf\footnotesize
\begin{tabular}{@{}c@{\;\;\;}c@{\;\;\;}c@{\;\;\;}c@{\;\;\;}c@{\;\;\;}l@{\;\;\;}r@{\;\;\;}r@{\;\;\;}c@{\;\;\;}l@{}}
\toprule
region & trait & CS & size & purity & sentinel SNP & {\em p-}value & PIP &
alleles* & candidate gene(s) \\
\midrule
\multirow[t]{2}{*}{1q21.3} & \multirow[t]{2}{*}{COA} &
L1 & 1 & -- & rs61816761 & 1.95$\times$10\textsuperscript{-46} & >0.99 &
G/A & {\em FLG}, {\em FLG2}, {\em HRNR} \\ 
& & L2 & 2 & 0.850 & rs12123821 & 1.35$\times$10\textsuperscript{-33} &
0.93 & C/T & {\em FLG}, {\em HRNR}, {\em RPTN} \\ 
\multirow[t]{2}{*}{2q12.1} & \multirow[t]{2}{*}{AA} & L1 & 18 & 0.986 &
rs72823641 & 4.73$\times$10\textsuperscript{-42} & 0.47 &
A/T & {\em IL1R1}, {\em IL18R1}, {\em IL1RL2} \\ 
& & L2 & 61 & 0.992 & rs10179654 & 3.01$\times$10\textsuperscript{-02} &
0.06 & G/T & {\em IL1R1}, {\em IL18R1}, {\em IL1RL2} \\
2q22.3 & AOA & L1 & 32 & 0.933 & rs7571606 &
2.84$\times$10\textsuperscript{-09} & 0.05 & 
A/T & {\em TEX41}, {\em ACVR2A} \\ 
\multirow[t]{4}{*}{10p14} & \multirow[t]{4}{*}{AA} & L1 & 2 & 0.887 &
rs11256016 & 7.63$\times$10\textsuperscript{-19} & 0.84 & G/A & {\em GATA3} \\
& & L2 & 2 & 0.970 & rs72782675 & 6.10$\times$10\textsuperscript{-11} &
0.77 & T/C & {\em GATA3} \\ 
& & L3 & 3 & 0.980 & rs12413578 & 4.27$\times$10\textsuperscript{-26} &
0.64 & T/C & {\em GATA3} \\ 
& & L4 & 10 & 0.996 & rs2197415 & 2.47$\times$10\textsuperscript{-36} &
0.23 & G/T & {\em GATA3} \\
\multirow[t]{2}{*}{11q13.5} & \multirow[t]{2}{*}{AA} 
& L1 & 2 & 0.943 & rs55646091 & 5.11$\times$10\textsuperscript{-25} &
0.61 & G/A & {\em EMSY}, {\em LRRC32}, {\em THAP12} \\
& & L2 & 10 & 0.900 & rs7936323 & 1.07$\times$10\textsuperscript{-37} &
0.18 & G/A & {\em EMSY}, {\em LRRC32}, {\em THAP12} \\
12q13.11 & AOA & L1 & 9 & 0.909 & rs56389811 &
5.36$\times$10\textsuperscript{-11} & 0.22 & T/C
& {\em HDAC7}, {\em SLC48A1} \\
15q22.2 & AA & L1 & 18 & 0.770 & rs11071559 &
4.23$\times$10\textsuperscript{-13} & 0.25 & T/C & 
{\em RORA}, {\em ANXA2} \\ 
17q12 & COA & L1 & 8 & 0.999 & rs4795399 & 1.97$\times$10\textsuperscript{-85}
& 0.19 & C/T & {\em GSDMB}, {\em ORMDL3} \\ \bottomrule
\end{tabular}
\vspace{1ex}
\caption{\rm Summary of the CoxPH-SuSiE asthma fine-mapping results.
  The columns in the table from left to right are: asthma locus; trait
  analyzed (COA = childhood-onset asthma; AOA = adult-onset asthma; AA
  = all asthma); the credible set (CS) label used in
  Figure~\ref{fig:susie1}; number of SNPs in the 95\% CS; {\em purity},
  defined as the smallest absolute correlation (Pearson's {\em r})
  among all SNP pairs in the CS \citep{wang2020simple}; the sentinel
  SNP (the SNP with the largest PIP in the CS); the SPACox association
  test {\em p}-value for the sentinel PIP; the PIP of the sentinel
  SNP; and ``candidate genes'', a non-comprehensive listing of
  candidate asthma genes based on the fine-mapping results and
  previous asthma GWAS.
  %
  %
  More comprehensive and detailed results are provided in the
  Supplementary Data. *Alleles are reported
  as A/B, where B is the allele that increases the hazard function.
\label{tab:susie}}
\end{table}

The fine-mapping results are summarized in Figure \ref{fig:susie1} and
Table \ref{tab:susie}; see the Supplementary Data
for more detailed results. At 4 of the 8
asthma loci, CoxPH-SuSiE identified multiple CSs, suggesting the
presence of multiple causal variants. At one locus, 10p14, CoxPH-SuSiE
identified 4 CSs, suggesting the presence of 4 distinct asthma risk
variants. The ability of fine-mapping analyses to identify multiple
causal variants underlying an association signal is an important
motivation for fine-mapping.

The strongest associations (Table~\ref{tab:asthma_loci}) are often
expected to show up as SNPs with large PIPs in a CS. Consistent with
this, all but one of the top SNPs were also the sentinel SNP in a CS
(Table \ref{tab:susie}), and the other top SNP (rs11236797, at
11q13.5) was one of 10 SNPs included in a CS for that region.

The analysis of asthma locus 2q12.1 identified a CS containing SNPs
with very little evidence for marginal association
(Figure~\ref{fig:susie1}). This can happen in a fine-mapping analysis
when a signal becomes apparent only after accounting for---i.e.,
conditioning on---the effects of other causal SNPs. Indeed, the
potential to identify such ``secondary'' signals is another reason to
conduct a fine-mapping analysis.
To check that this was the explanation here, we performed a
``conditional'' association analysis for the SNPs at 2q12.1:
we conditioned on the genotype of rs72823641 (the top SNP in the first
CS) by including it as an additional covariate in the CoxPH
model. This association analysis, shown in Figure \ref{fig:cond_pval},
produced much more significant {\em p-}values for the SNPs in the
second CS, consistent with the fine-mapping results that suggest the
presence of 2 causal SNPs at this locus.

Among the CSs returned by CoxPH-SuSiE, 6 included a SNP with a PIP of
0.5 or greater (Table \ref{tab:susie}, Supplementary Data). These SNPs
are of particular interest because they are likely to be the SNPs that
affect asthma risk, or they might tag other genetic variants
(unavailable to us) that are the asthma risk variants. For comparison,
\cite{zhong2025integration}, using SuSiE with summary statistics to
finemap asthma risk loci, obtained PIP $>$ 0.5 for 3 of these 6
SNPs. They also identified rs12413578 with PIP = 0.44. But two of the
causal SNPs we identified the 10p14 locus, rs11256016 and rs72782675,
were not included in any of their CSs.

The two SNPs with the highest PIPs were both in the 1q21.3 locus. Both
SNPs have been previously reported as pathogenic or tagging pathogenic
variants: rs61816761 (PIP $>$ 0.999 for COA and AA) is a stop-gain
mutation (c.16819G>A) that occurs in an exon of the gene {\em FLG}
(\citealt{smith2006loss, palmer2006common}; ClinVar accession
VCV000016319.70, OMIM entry 135940); rs12123821 (PIP $>$ 0.93 for COA
and AA) tags a loss-of-function mutation in {\em FLG}, c.2282del4
(\citealt{marenholz2017serpinb}; OMIM entry 135940). Our fine-mapping
results further support these two variants as affecting asthma risk.

The other 4 SNPs with a PIP of 0.5 or greater are all intergenic,
non-coding variants for which their regulatory function and
pathogenicity are unknown. One SNP, rs55646091 (PIP = 0.61), is not
the top association at the 11q13.5 locus, and therefore may have been
missed by some previous studies (the top association, rs11236797, is
in the other CS at that locus).  Interestingly,
\cite{tomizuka2024predicted} also identified rs55646091 as a strong
putative causal variant for asthma, although did not attribute a
regulatory function to this variant. The three
remaining SNPs with PIP $>$ 0.5 are all in the 10p14
locus. \cite{zhong2025integration} This locus is interesting because
many immune-related traits map to this locus, including rheumatoid
arthritis, multiple sclerosis, type 1 diabetes and allergic diseases.
Interpreting these genetic associations is challenging because the
associated variants are far from any protein coding sequence. However,
a very recent study of {\em GATA3}, which is the closest gene to these
putative causal variants, discovered a 44-kb regulatory sequence
located approximately 1 Mb downstream of {\em GATA3} that is a distal
enhancer of {\em GATA3} in Th2 cells \citep{chen2023deletion}. All
three candidate causal SNPs we identified at this locus lie at
base-pair positions within this 44-kb regulatory sequence.

For completeness, we ran all three types of fine-mapping analysis
(COA, AOA, AA) on each region (Supplementary Data). The results were
consistent with expectations from the marginal associations: in
regions identified as COA only (e.g., 1q21.3) CSs were identified in
the COA fine-mapping analysis but not in AOA analysis; and in regions
identified as both COA and AOA (e.g., 10p14) more CSs were sometimes
identified in the combined AA analysis than either COA or AOA alone,
highlighting the benefits of a combined analysis in these cases.

\section{Discussion}
\label{sec:coxph_susie_discussion}


In this paper, we presented CoxPH-SuSiE, a new Bayesian variable
selection method for censored TTE phenotypes, and illustrated its
potential on both simulated and real data.

A core computation in CoxPH-SuSiE is the computation of the Bayes
factor for the single-variable CoxPH regression model.  This
computation is much harder than the corresponding computation for the
linear model in SuSiE (where the Bayes factor is available in closed
form).  The algorithm for fitting CoxPH-SuSiE (Algorithm
\ref{algo:gibss}) performs this computation many times; specifically,
$pL$ times in each iteration of the repeat-until loop, where $p$ is
the number of covariates, and $L$ is the number of SERs. To help
reduce computation, we proposed an approximate Bayes factor using a
variant of Laplace's method, and in experiments we found that this
approximate Bayes factor achieved a good balance of accuracy and
speed. This approximation makes CoxPH-SuSiE practical for data sets
containing hundreds of thousands of samples and over a thousand
covariates.

We also exploited the fact that these Bayes factor computations can
easily be performed in parallel to reduce the running times. For
example, when we applied CoxPH-SuSiE to the 10p14 asthma locus ($n =
\mbox{268,829}$ individuals, $p = \mbox{1,651}$ SNPs), running on a
machine with a single processor took about 6 hours, whereas running on
a multicore machine with 10 processors reduced the running time to
just over 1 hour. (Details on the computing environment used to obtain
these running times are given in the Appendix.)  There is potential to
further speed up CoxPH-SuSiE by exploiting fast implementations of the
CoxPH model in R packages such as SPACox \citep{bi2020fast}, COXMEG
\citep{coxmeg} or Colossus \citep{colossus} (See
\citealt{cran_task_view} for a survey of R packages implementing the
CoxPH model.) We used the survival R package \citep{survival-book} in
our implementation because it is a robust and widely used software
package, but potentially any CoxPH software implementation could be
used provided that it gives the three statistics needed to compute the
Laplace Bayes factor (eq.~\ref{eq:BF_lap}): the maximum-likelihood
estimate of $b$, its variance, and the likelihood ratio.



There are also practical issues that one should consider when using
CoxPH-SuSiE. First, it is important to consider whether the
``proportional hazards'' assumption is reasonable.
%
%
In our asthma fine-mapping analyses, we divided asthma cases into
childhood-onset asthma and adult-onset asthma using the
age thresholds from \cite{pividori2019shared}. However, these
thresholds are somewhat arbitrary, and a more systematic choice of the
age thresholds could itself be an interesting problem in the study of
asthma subtypes. For TTE data where the proportional hazards
assumption does not hold, a better approach might be to incorporate
time-varying effects into the CoxPH model
\citep{ojavee2023genetic}. Other  TTE models that have been used in
genome-wide association studies include frailty models
\citep{dey2022efficient} and age-dependent liability models
\citep{adult}.
%
%
Potentially, these models could also be combined with the SuSiE prior
to create alternative fine-mapping methods for TTE phenotypes.

Second, we introduced a model-fitting procedure which we called
``Generalized Iterative Bayesian Stepwise Selection'', or ``gIBSS''
for short (Algorithm \ref{algo:gibss}). This algorithm generalizes the
IBSS algorithm for SuSiE \citep{wang2020simple}. However, unlike IBSS,
there is no guarantee that gIBSS will converge to a stationary point
of a specific objective function. Deriving a
model-fitting procedure that can be understood as optimizing an
approximate posterior---that is, a variational inference method
\citep{blei2017variational}---and understanding the exact form of the
approximate posterior are open research questions.


\begin{acks}[Acknowledgments]
We thank
%
%
Yuxin Zou, Brandon Pierce and Xin He for their help with the
processing and analysis of the UK Biobank data. We thank Gao Wang and
Yuxin Zou for their help with the GTEx data, and for developing the
susieR software. Gao Wang also developed and supported the Dynamic
Statistical Comparisons (DSC) software used to implement the
simulations. We also thank the staff at the Research Computing Center
(RCC) and the staff the Center for Research Informatics (CRI) at the
University of Chicago for maintaining the high-performance computing
resources used to run the numerical experiments. This research has
been conducted using data from UK Biobank, a major biomedical database
(UK Biobank Application Number 27386).
\end{acks}

\begin{funding}
This work was supported by the NHGRI at the National
Institutes of Health under award number R01HG002585.
\end{funding}

\begin{supplement}
\stitle{Supplementary Data}
\sdescription{Table in CSV format giving the results of the
  CoxPH-SuSiE asthma fine-mapping analyses.}
\end{supplement}


\bibliographystyle{imsart-nameyear}
\bibliography{coxph_susie}

\renewcommand\figurename{Supplementary~Figure}
\setcounter{figure}{0}

\appendix

\section{Derivations}

This section contains derivations supporting results in the main text.

\subsection{EM for estimating the prior variance in the SER CoxPH model}
\label{appendix_em}

Here we derive the EM update \eqref{eq:sigma0-em-update} for
estimating the prior variance in the SER CoxPH model (defined in
Section \ref{sec:coxph_ser} of the main text).

To derive the EM update, we use the multinomial random variable
${\bm\gamma}$ for the data augmentation. With this data augmentation,
the complete data log-likelihood is
\begin{equation}
\mathscr{L}(\sigma_0^2) =
\sum_{j=1}^p \delta_1(\gamma_j) \log p(b_j \mid \sigma_0^2),
\end{equation}
where $\delta_y(x)$ denotes the Dirac ``delta'' function that is 1 at
$x = y$, and zero at any $x \neq 0$.

where ``constant'' absorbs the additional terms in the log-likelihood
that do not depend on $\sigma_0^2$.

The M-step is derived by solving for the $\sigma_0^2$ that maximizes
the expected complete log-likelihood. As with the other calculations
for this model, we maximize an approximate expected completed
log-likelihood,
\begin{align}
E[\mathscr{L}(\sigma_0^2)] &=
-\frac{1}{2} \log\sigma_0^2 - \frac{1}{2\sigma_0^2}
\sum_{j=1}^p E[b_j^2 \mid \gamma_j = 1] \, p(\gamma_j = 1) +
\mbox{constant} \nonumber \\
&\approx
-\frac{1}{2}\log\sigma_0^2
- \frac{\sum_{j=1}^p\alpha_j(\mu_{1j}^2 + \sigma_{1j}^2)}{2\sigma_0^2}
+ \mbox{constant.}
\end{align}
Taking the derivative of this approximate expected complete
log-likelihood, setting it to zero, and solving for $\sigma_0^2$
results in the closed-form solution \eqref{eq:sigma0-em-update}.

\subsection{Exponential distribution property}
\label{supple:sim_survT}

Let $X$ and $Y$ be exponentially distributed with rates $\lambda$ and
$\mu$, respectively. Then $\Pr(X > Y) = \frac{\mu}{\mu+ \lambda}$.

\begin{proof}
\begin{align}
\Pr(X > Y) &= \int_0^{\infty}
\int_0^{x}\lambda e^{-\lambda x} \mu e^{-\mu y} \, dy \, dx \nonumber \\
&= \int_0^{\infty}\lambda e^{-\lambda x} \, dx -
\int_0^{\infty}\lambda e^{-\lambda x - \mu x} \, dx \nonumber \\
&= 1 - \frac{\lambda}{\lambda + \mu} \int_0^{\infty}(\lambda+\mu)
e^{-(\lambda + \mu) x} \, dx\nonumber \\
&= \frac{\mu}{\lambda + \mu}.
\end{align}
\end{proof}

\section{Computing the Bayes factors using Gaussian-Hermite quadrature}
\label{app:GHQ}

Here we briefly describe the numerical integration technique,
Gauss-Hermite quadrature, that was used to accurately estimate the
Bayes factors \eqref{eq:BF_lap} for the simple single-variable
CoxPH regression model \eqref{eq:coxph}.

\cite{naylor1982applications} give the
following approximation formula (eq.~9 in their paper):
\begin{equation} 
\int_{-\infty}^{\infty} f(t) \, \phi(t; \mu, \sigma^2) \, dt \approx
\frac{1}{\sqrt{\pi}} \sum_{i=1}^{n} w_i f(t_i),
\label{eq:gauss_hermite_reparam}
\end{equation}
where $\phi(x; \mu, \sigma^2)$ denotes the probability density
function of the normal distribution with mean $\mu$ and variance
$\sigma^2$ at $x$, and $t_i \colonequals \mu + \sqrt{2} \sigma r_i$,
in which the $r_i$ are the zeros of the $n$th-order Hermite
polynomial, with corresponding weights $w_i$. See also
\cite{davis2007methods}. This formula can be used to approximate an
integral $I = \int_{-\infty}^\infty g(t) \, dt$ by expressing it as
\begin{equation}
I = \int_{-\infty}^{\infty} g(t) \, dt =
\int_{-\infty}^{\infty} h(t) \, \phi(t; \mu, \sigma^2) \, dt  
\end{equation}
where $h(t) \colonequals g(t )/ \phi(t;\mu, \sigma^2)$. Here, $\mu$
and $\sigma^2$ are arbitrary, and so should be chosen to make the
approximation as accurate as possible. Results in \cite{liu1994note}
suggest that $\mu, \sigma^2$ should be chosen so that $g(x) \approx
\alpha \phi(x; \mu, \sigma^2)$ for some constant $\alpha$.

The numerator in the BF \eqref{eq:BFp} is an integral, $I$, with $g(t)
= \ell(t; {\bm x}, {\bm c}) \, \phi(t; 0, \sigma_0^2)$, which is
proportional to the posterior distribution of $b$ \eqref{eq:post}. We
therefore select $\mu$ and $\sigma^2$ to be the approximate posterior
mean \eqref{eq:mu1} and approximate posterior variance
\eqref{eq:sigma1}, then apply the approximation formula
\eqref{eq:gauss_hermite_reparam} with $n = 32$.

Note that the special case of \eqref{eq:gauss_hermite_reparam} with $n
= 1$ recovers the Laplace BF \eqref{eq:BF_lap}. Larger values of $n$
will provide increasingly accurate approximations of the BF.
%
%
See \cite{liu1994note} for more discussion on the relationship between
Laplace's method and Gauss-Hermite quadrature.

\section{Preparation of the data for the simulations and asthma
  fine-mapping analyses}
\label{sec:data-prep}

\subsection{GTEx data for simulations}

We used genotype data from release v7 of the GTEx Project
\citep{gtex2015genotype, gtex2017} to simulate fine-mapping data sets.
We arbitrarily chose a region on chromosome 1 near gene {\em ANGPTL3}
as the candidate fine-mapping region. After removing SNPs with minor
allele frequencies (MAFs) less than 1\%, the GTEx genotype data
consisted of genotypes from 574 individuals at 7,154 SNPs.

\subsection{UK Biobank data for simulations}


We used version 3 of the imputed UK Biobank genotypes
\citep{bycroft2018uk} to simulate fine-mapping data sets. We
arbitrarily chose a region on chromosome 3 near gene {\em SATB1-AS1}
as the candidate fine-mapping region. After removing SNPs with MAFs
less than 0.1\%, the UK Biobank genotype data consisted of genotypes
from 248,980 individuals at 3,119 SNPs. The exact steps taken to
prepare these genotype data, including steps to remove individuals
based on certain criteria, are described in \cite{zou2023fast}.

For the simulations, SNPs with MAFs less than 1\% were also removed,
then further subsampled, separately for each simulation, to obtain
1,000 SNPs. Also, the individuals were subsampled at random,
separately for each simulation, to obtain a random set of 50,000
individuals.

\subsection{UK Biobank data for asthma association and fine-mapping analyses}


We used a subset of available UK Biobank genotypes (version 3),
removing samples that met one or more of the following criteria for
exclusion (similar to the exclusion criteria used in
\citealt{zou2023fast}): mismatch between self-reported and genetic
sex; withdrew from UK Biobank; didn't know, or didn't wish to answer,
their asthma diagnosis age. Additionally, we excluded outlying
genotype samples based on heterozygosity and/or rate of missing
genotypes as defined by UK Biobank (data field 22027), and we removed
any individuals having at least one relative in the cohort based on UK
Biobank kinship calculations (samples with a value other than zero in
data field 22021). Finally, to limit confounding due to population
structure, we included only genotype samples marked as ``White
British'' (stored in data field 22009, and based on a PCA of the
genotypes; \citealt{bycroft2018uk}). After filtering genotype samples
according to these criteria, 273,543 samples remained.

Censored/uncensored status for the three asthma phenotypes was defined
as follows. For AA, an individual was uncensored if a doctor diagnosis
for asthma was reported (data fields 3786, 22147); a censored
individual was any other individual.  Using the same data fields, COA
uncensored individuals were individuals who developed asthma at or
before age 12, and censored were individuals who developed asthma at
or after age 26, or did not report an asthma diagnosis. AOA uncensored
individuals were defined as individuals who developed asthma between
(and including) ages 26 and 65, and controls were individuals who did
not report an asthma diagnosis during the course of the study. These
phenotype definitions come from \cite{pividori2019shared}.  For
uncensored individuals, the event time was defined as the age of
diagnosis (in years) as described above. For censored individuals, the
censoring times were defined as follows: for AA, the age at the most
recent visit to the assessment center (data field 21003); for COA,
always 12 years; for AOA, 65 years or current age, whichever was
smaller.  Table \ref{tab:asthma_data} in the main text summarizes the
TTE data for the three asthma phenotypes.

The following covariates were included in the SPACox association
analyses and the CoxPH-SuSiE fine-mapping analyses: sex (UK Biobank
data field 31) and genetic PCs 1--10 (data field 22009).

%
%

%
%

%
%

\section{Details of the simulations}
\label{sec:data_generation}

%
%
Here we describe the steps that were taken to simulate the
fine-mapping data sets.

Simulating censored TTE data requires three inputs: an $n \times p$
design matrix, $\X$; $p_1 \leq p$, the number of effect variables; the
mean $\mu$ and variance $\sigma^2$ for simulating the coefficients of
the effect variables; and the censoring level, $r \in [0, 1)$. In the
comparisons of the different BVSR methods, $\X$ was always a genotype
data matrix from a real data set. In the Bayes factor comparisons,
$\X$ was an $n \times 1$ matrix containing the genotypes for a single
simulated SNP, $x_i \sim \mathrm{Binom}(2,f)$, where $f \in (0, 0.5]$
is the SNP minor allele frequency.

The following simulation procedure generates a vector of event times,
$\y = (y_1, \ldots, y_n)^{\intercal}$, and a vector giving the
censoring status for each sample, ${\bm\delta} = (\delta_1, \ldots,
\delta_n)^{\intercal}$:
  

%
\begin{enumerate}

\item Randomly sample without replacement the $p_1 \leq p$ indices of
  the effect variables from $\{1, \ldots, p\}$.

\item Simulate the vector of coefficients, $\b = (b_1, \ldots,
  b_p)^{\intercal}$: if $j \in \{1, \ldots, p\}$ is one of the indices
  selected in the first step, $b_j \sim N(\mu, \sigma^2)$; otherwise,
  $b_j = 0$. Note that in the Bayes factor comparisons, we set
  $\sigma^2 = 0$ so that all $b_j$ were either 0 or $\mu$, with $\mu =
  0.1$.

\item For each sample $i = 1, \ldots, n$, compute the survival rate
  as $\lambda_i^s = b_0 + \x_i^{\intercal} \b$. In all
  our simulations, we set $b_0 = 1$.

\item For each sample $i = 1, \ldots, n$, simulate the survival time
  and censor time as $T_i \sim \mathrm{Expon}(\lambda_i^s)$ and $C_i
  \sim \mathrm{Expon}(\lambda^c)$, where $\lambda^c =
  \frac{r\bar{\lambda}^s}{1 - r}$, and $\bar{\lambda}^s =
  \frac{1}{n}\sum_{i=1}^n\lambda^s_i$.  See below for discussion on
  the formula for $\lambda^c$.
        
%
%

\item Finally, record the event time and censoring status: $y_i =
  \min(T_i, C_i)$ and $\delta_i$ is 1 if $T_i \le C_i$, otherwise it
  is zero.
  
\end{enumerate}
%
%
Note that data simulated in this way satisfies the proportional hazard
assumption.

Regarding the censoring rate, $\lambda^c$, we would ideally like
$\lambda^c$ to satisfy
\begin{align}
r = \frac{1}{n}E\big[{\textstyle \sum_i^n\mathbb{I}(C_i < T_i)}]
= \frac{1}{n} \sum_{i=1}^n \Pr(C_i < T_i)
= \frac{1}{n} \sum_{i=1}^n \frac{\lambda^c}{\lambda^s_i + \lambda^c},
\label{eq:solve_censor_rate}
\end{align}
in which the last equality uses the result from Section
\ref{supple:sim_survT}. Therefore, the choice $\lambda^c =
\frac{r\bar{\lambda}^s}{1 - r}$ is an approximation to this ideal
choice intended to simplify the simulation procedure.

\section{Computing environment}

The simulations were performed using R 4.2.0 \citep{R}, linked to the
OpenBLAS 0.3.13 optimized numerical libraries, on Linux machines
(Scientific Linux 7.4) with Intel Xeon E5-2680v4 (``Broadwell'')
processors. The following R packages were used: BVSNLP 1.1.9, R2BGLiMS
0.1-07-02-2020, survival 3.3-1, susieR 0.12.35, survival.svb 0.0-2 and
statmod 1.5.0. All computations in the simulations were performed
using a single processor.

The asthma fine-mapping analyses were performed using R 4.3.1, linked
to the OpenBLAS 0.3.20 optimized numerical libraries, on Linux
Machines with Intel Xeon Gold 6326 processors. The following R
packages were used: susieR X, survival X and SPACox X. Also,
PLINK v2.00a2LM 64-bit Intel (21 Feb 2019) \citep{chang2015second} was
used to prepare the UK Biobank genotype data (Section
\ref{sec:data-prep}).
 

\newpage

\begin{figure}[t]
\centering
\includegraphics[width=0.9\textwidth]{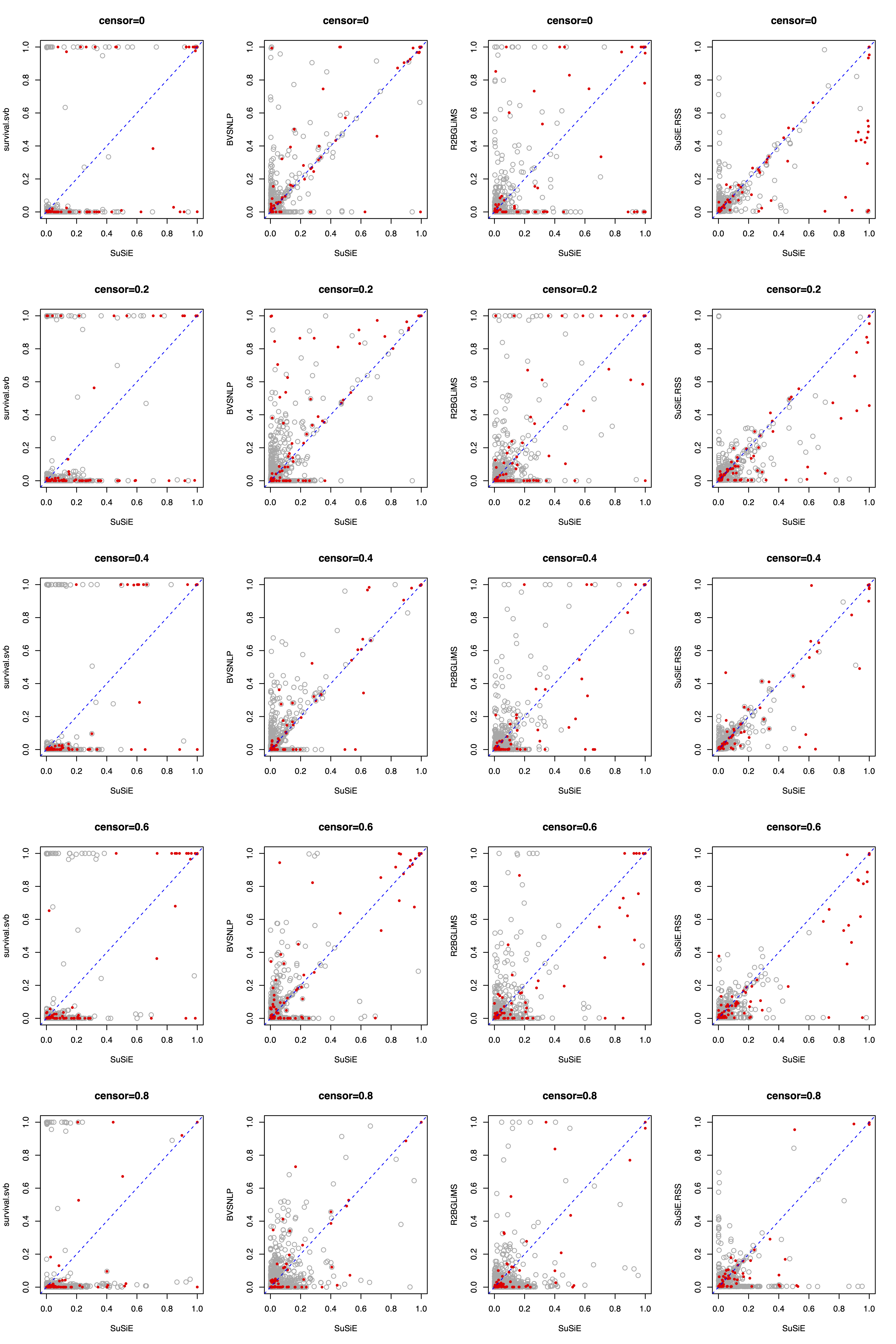}
\caption{\rm CoxPH-SuSiE PIPs ($x$-axis) vs. PIPs from other methods
  ($y$-axis) in the GTEx simulations, separately at different
  censoring levels. Each point is single SNP; causal SNPs are shown as
  solid red circles, and other SNPs are shown as open gray circles.}
\label{fig:pip_gtex}
\end{figure}

\begin{figure}[t]
\centering
\includegraphics[width=0.85\textwidth]{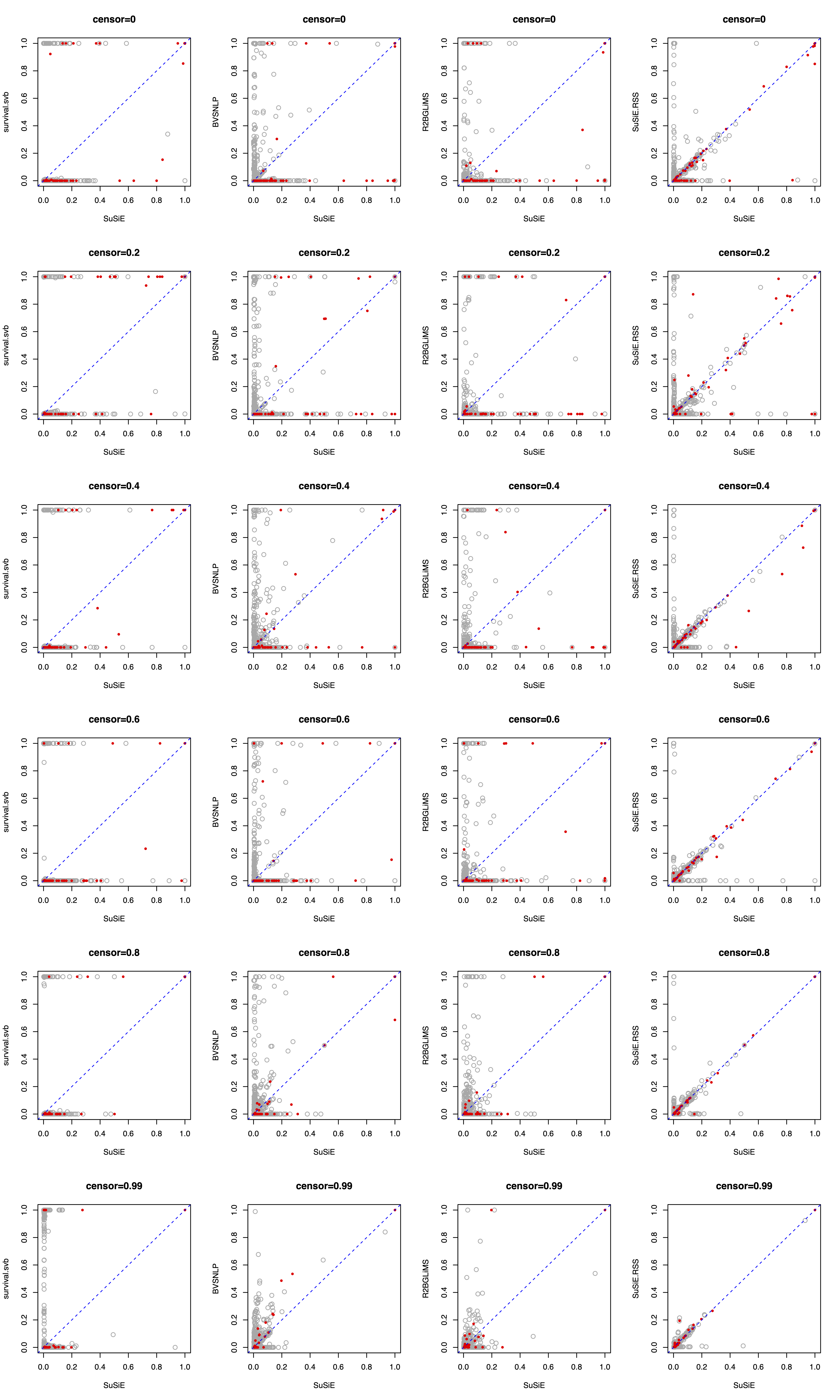}
\caption{\rm CoxPH-SuSiE PIPs ($x$-axis) vs. PIPs from other methods
  ($y$-axis) in the UK Biobank simulations, separately at different
  censoring levels. Each point is a single SNP; causal SNPs are shown
  as solid red circles, and other SNPs are shown as open gray
  circles.}
\label{fig:pip_ukb}
\end{figure}

\begin{figure}[t]
\centering
\includegraphics[width=\textwidth]{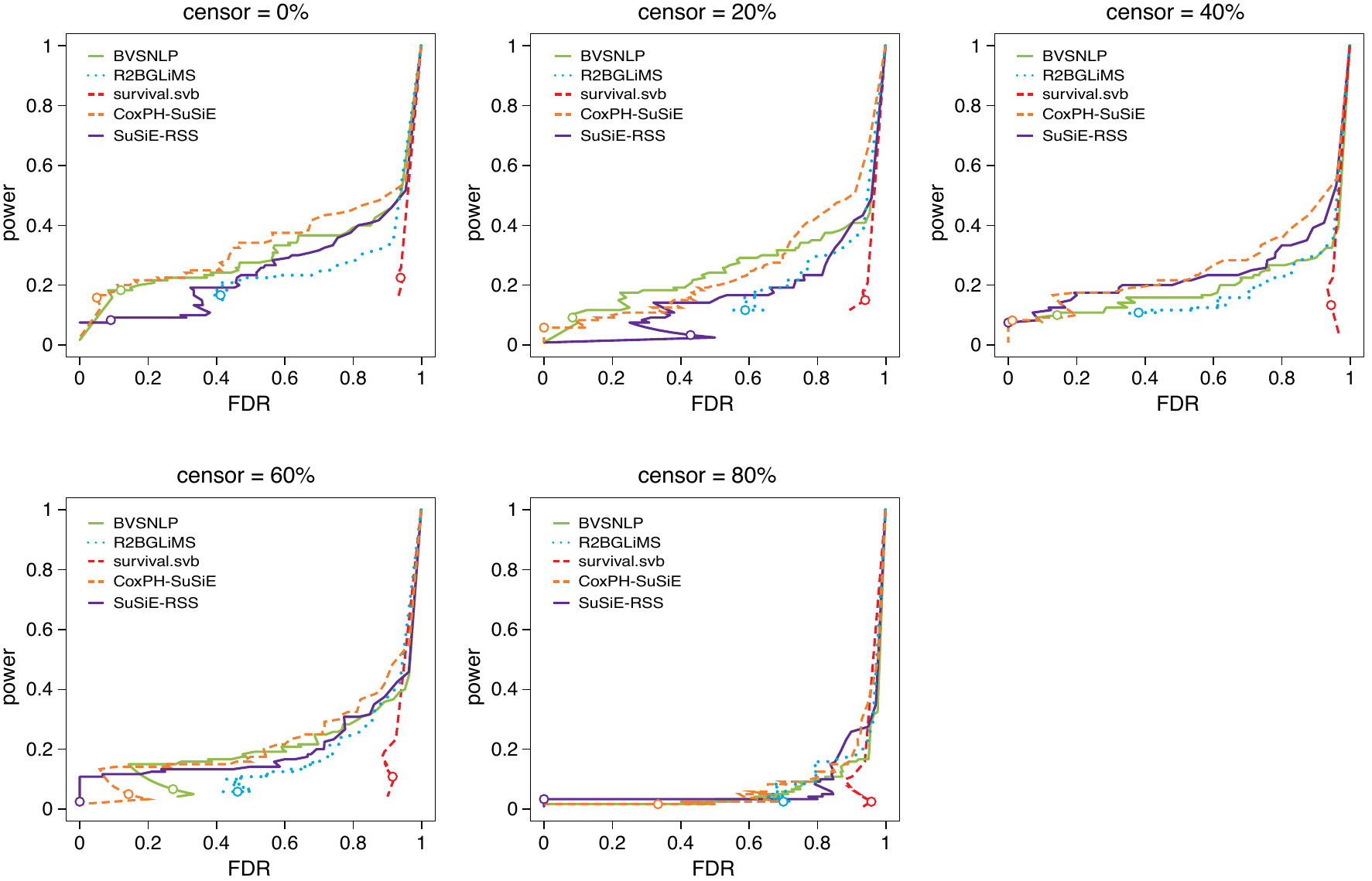}
\caption{\rm Discovery of causal SNPs using PIPs in the GTEx
  simulations. Each curve shows power vs. FDR in identifying causal
  SNPs. For each power-vs-FDR plot, the FDR and power were calculated
  from the results on 80 data sets as the PIP threshold was varied
  from 0 to 1. Open circles are drawn at a PIP threshold of 0.95.}
\label{fig:fdr_gtex}
\end{figure}

\begin{figure}[t]
\centering
\includegraphics[width=\textwidth]{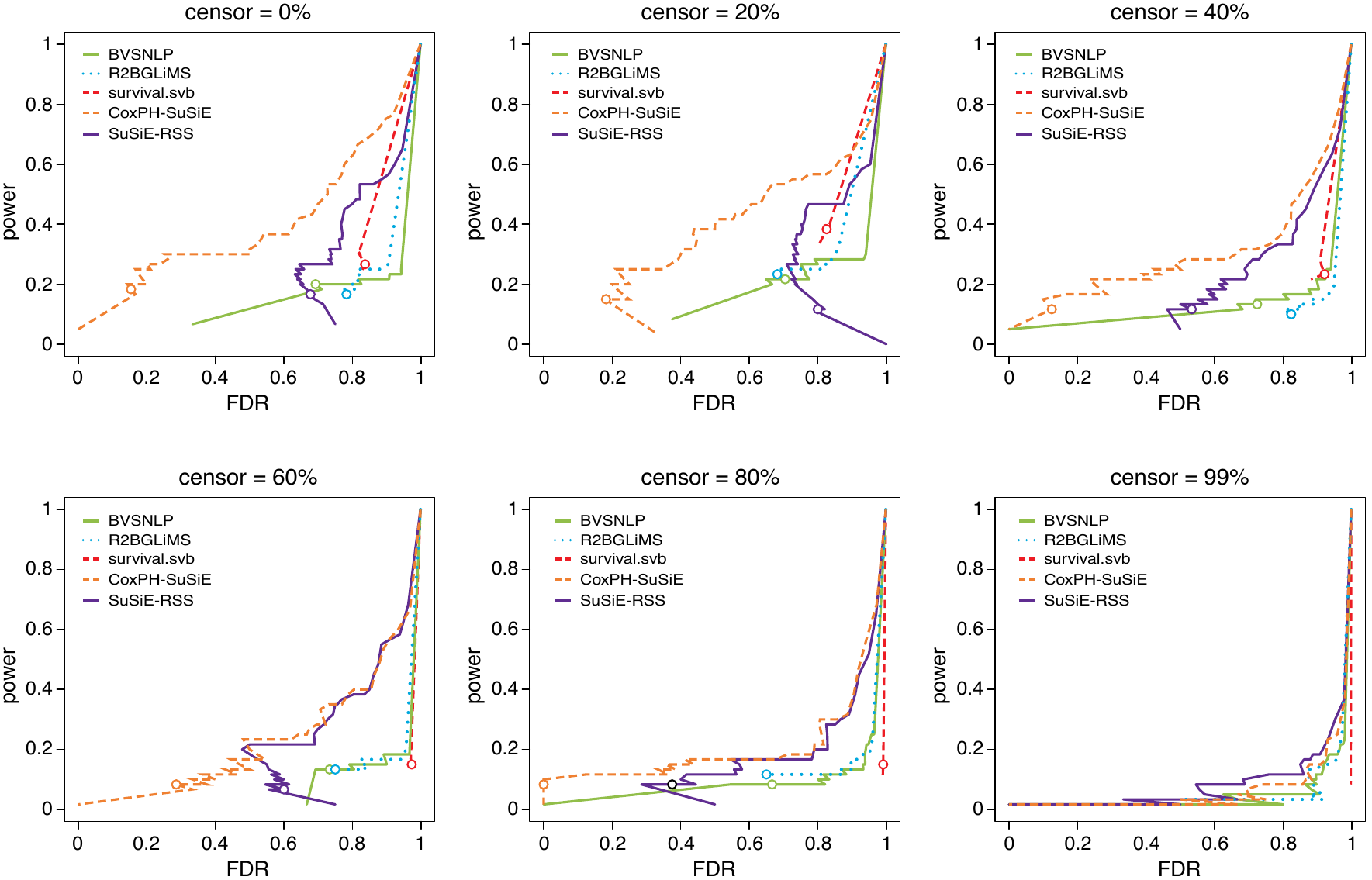}
\caption{\rm Discovery of causal SNPs using PIPs in the UK Biobank
  simulations. Each curve shows power vs. FDR in identifying causal
  SNPs. For each power-vs-FDR plot, FDR and power were calculated from
  the results on 40 data sets as the PIP threshold was varied from 0
  to 1. Open circles are drawn at a PIP threshold of 0.95.}
\label{fig:fdr_ukb}
\end{figure}

\end{document}